\documentclass[journal]{IEEEtran}
\usepackage{ifpdf}
\usepackage{cite}
\usepackage[utf8]{inputenc}
\usepackage{booktabs}
\usepackage{longtable}
\usepackage[table]{xcolor}
\usepackage{colortbl}
\usepackage{hhline}
\usepackage{multicol}
\usepackage{url}
\ifCLASSINFOpdf
\usepackage[pdftex]{graphicx}
\graphicspath{{../pdf/}{../jpeg/}}
\DeclareGraphicsExtensions{.pdf,.jpeg,.png}
\fi
\usepackage{amsmath}
\usepackage{algorithmic}
\usepackage{array}
\usepackage{multicol}
\usepackage{lipsum}
\usepackage{color}
\definecolor{lighterheadercolor}{RGB}{252, 213, 180}
\definecolor{black}{RGB}{0,0,0}
\usepackage{stfloats}
\usepackage{url}
\begin{document}
\title{Bacterial Communications and Computing in Internet of Everything (IoE)}

\author{B. Yagmur Koca, \IEEEmembership{Student Member,~IEEE}, and Ozgur B.~Akan, \IEEEmembership{Fellow,~IEEE}
\thanks{The authors are with the Center for neXt-generation Communications (CXC), Department of Electrical and Electronics Engineering, Ko\c{c} University, Istanbul 34450, Turkey (e-mail: yakoca23@ku.edu.tr).}
\thanks{O. B. Akan is also with the Internet of Everything (IoE) Group, Electrical Engineering Division, Department of Engineering, University of Cambridge, Cambridge CB3 0FA, UK (email: oba21@cam.ac.uk).}
\thanks{This work was supported in part by the AXA Research Fund (AXA Chair for Internet of Everything at Ko\c{c} University)}
}
\maketitle
\begin{abstract}

Concurrent with advancements in molecular communication (MC), bacterial communication is emerging as a key area of interest. Given the frequent use of bacteria in various MC models, it is essential to have a thorough grasp of their intrinsic communication, signaling, and engineering techniques. Although it is crucial to have a strong understanding of the communication background, the inherent biological variability of bacteria may introduce complexity. Thus, an in-depth understanding of bacteria and their communication is a must for improving and extending the models in which they are utilized. Furthermore, the emerging and evolving domain of bacterial computing provides an exciting opportunity for advancing applications in areas such as environmental monitoring and biological computing networks. By integrating the communication and sensing capabilities, bacterial computing offers a promising framework for enhancing the adaptability and responsiveness of bacteria. This intersection of bacterial sensing, communication and computing technologies may bring new possibilities for improving existing models and exploring novel applications.

This paper provides a comprehensive review of bacterial communication and computing, illustrating their application and the link with the concept of the Internet of Everything (IoE). Through the analysis of these biological systems, we reach a deeper insight on how the small-scale interactions may contribute to the major concept of universal interconnectedness; thus, we make the knowledge to flow and communication stronger between different fields. The discussion include the identification of the different bacterial mechanisms that could revolutionize the traditional communication systems. These discoveries underscore the remarkable ability of bacteria to adapt and respond through the integration of bacterial communication and sensing capabilities. Thus, this paper offers valuable insights into previously unaddressed aspects of bacterial behavior, suggesting novel avenues for future research and aiming to advance understanding and application of bacterial sensing, communication and computing in MC models. 

\end{abstract}
\IEEEpeerreviewmaketitle
\section{Introduction}
\IEEEPARstart{M}{olecular} communication (MC) is an emerging communication paradigm centered on the exchange of molecules to facilitate the transfer of information \cite{key3}. Due to its capacity for information exchange at the nano-scale within the human body, MC is considered the most favorable approach for nanonetworks and the Internet of Bio-Nano Things (IoBNT) \cite{key2,key124, key125, key126, key137}. In this context, bacterial communication is a rapidly growing area of research. Bacteria were previously presumed to be comprised of cells that have a limited number of functions governing the simple cellular processes, but their communication with the self and the external has gained importance over time \cite{key1}. Along with the breakthroughs in nanotechnology, it is very crucial to understand the biological mechanisms that support the communication system. Knowing these principles is as important for understanding, at a molecular level, the chemistry of signaling as it is for addressing communication challenges \cite{key37}.
 
The main focus of the studies in the literature that investigate MC is the modeling of the communication systems \cite{key124, key125, key204}. Although this method helps improve our understanding of the network, the efficiency and accuracy of the models are strongly dependent on the reactions of the biological entities that were included in the model. In order to ensure the utmost relevance of studies and models in the realm of MC, it is crucial to scrutinize the components one by one and then to methodically evaluate each of them. To develop more authentic models, including the bacteria or their byproducts, we need fundamental knowledge of the bacterial communication mechanisms. In addition, the similarity of these natural biological entities with man-made devices is gaining a lot of interest and also helping the field of research evolve into futuristic technologies. Therefore, there is a need to assess these systems within the framework of molecular communication and engineering.

\begin{figure*}[ht]
\includegraphics[width=\linewidth]{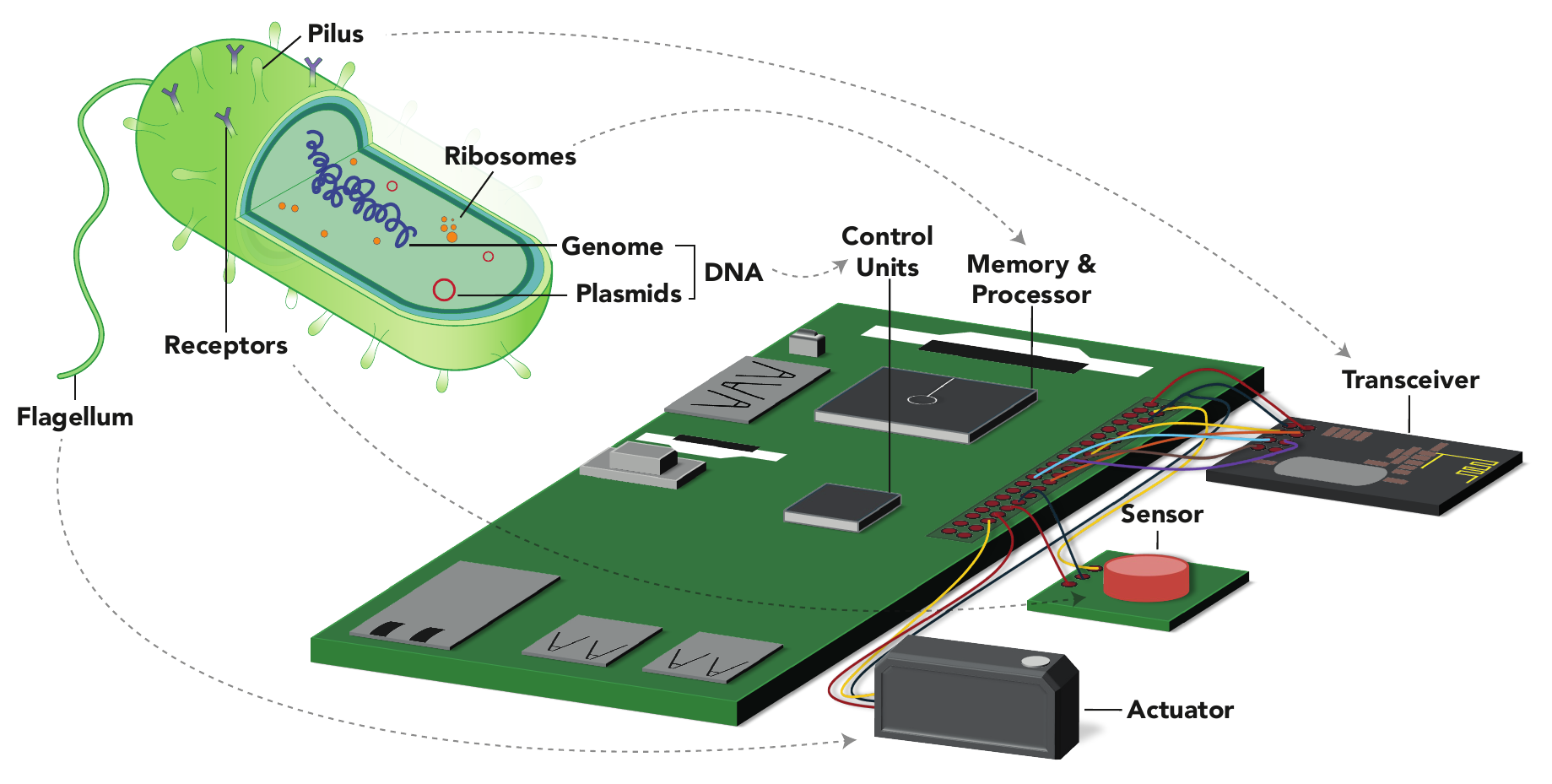}
\caption{Comparative illustration of bacterium and Internet of Things (IoT) device. This analogy underscores the functional parallels between biological entities and embedded systems \cite{key232}.}
\end{figure*}

As shown in Figure 1, the main components of the bacterium are depicted in a hybrid network of biologic and electronic structures of the Internet of Things (IoT) device. The pili of the bacterium serve as an anchor for the binding and transfer of the genetic material to other bacteria. It can be viewed as a transceiver, enabling communication and data transmission. Similarly, ribosomes, a part responsible for protein synthesis, work like memory units and processing units, which store and handle data, respectively. Sensing the changes in the environment through sensors in the bacteria serves the same function as sensors in an IoT device. The flagella of bacteria that they use in self-propulsion and steering can be compared to actuators that control or regulate a system. The DNA in bacteria at the level of the genome and plasmids controls the expressions of various functions and processes in a cell, which can also be analogous to a control unit in an IoT device.

Research on bacteria extends beyond communication models, including both theoretical and applied methods that utilize the inherent adaptability and intelligence. Studies have shown that bacteria have the capacity to communicate molecularly and evolve at a very rapid level with the ability to ‘learn’ molecularly \cite{key63}. From this perspective, the field of bacterial computing, both practical and very effective in nature \cite{key98, key99}, has not only provided solutions to problems found in current computer systems but has also tremendously helped conventional computing devices solve problems they cannot easily accomplish in a short amount of time \cite{key64, key65}. Bacterial computing per se is not a new sphere of exploration; however, focusing on the understanding and modeling of their communication paths has led to the development and modernization of this science. Bacterial computing differentiates itself from synthetically controlled communication technologies by using the intrinsic features of the bacteria's cells. These microorganisms have very simple yet manipulable structures; hence, they provide a unique opportunity for many modifications. Their participation in the sophisticated interrelationships positions them to be the greatest benefactor in several domains, implying that their presence in such a way gives them benefits through complex ecosystems.

Both of these concepts have been extensively researched, while answering many questions and also opening up new spaces for further developments. However, a number of problems still remain unexplored or to be resolved. Current MC models face many scalability and accuracy problems, especially when dealing with intricate biological processes and dynamic interactions. The gap between the theoretical models and their practical implementations presents a need for the development of more robust and agile MC systems. To establish these bacterial communication systems, an interdisciplinary approach is required. Examining the bacteria comprehensively as a collective entity and getting a good grasp of their interrelationships is important. Adopting an interdisciplinary frame for this subject will result in the MC models getting developed more accurately and will also enhance understanding related to the limitations that the subject holds. The approach to this field should not be limited to examining the processes. Besides, it is necessary to discover flaws and provide ideas for future studies. Therefore, this paper paves the way for a profound investigation into this emerging area.

In this work, we examine the existing body of literature to assess the contributions made in each of these topics and suggest potential areas for further research. Contribution of this review paper summarized as follows. 
\begin{itemize}
    \item The communication functionality of bacteria is reviewed, including sensing mechanisms and signaling molecules, with the intention of enriching our understanding of bacterial communication paradigms.
    \item Various bacterial communication methods are analyzed, offering a fresh perspective on how bacteria use diffusable molecules and electrochemical signals to communicate.
    \item Key methodologies of bacterial information transmission are reviewed, including existing research and some under-explored areas.
    \item Crucial bacterial characteristics that are pivotal for molecular communication research are identified and listed.
    \item An overview of advancements in bacterial computing, including bacteria-based molecular machine learning, quantum computing, and integrated systems, is presented.
    \item Potential future directions are highlighted, such as integrated sensing, communication, and computing in bacterial systems, thus paving the way for a new era of research on IoE.
\end{itemize}

\section{Quorum Sensing and Signaling Molecules}

Quorum sensing (QS) is an essential mechanism in bacterial communication, serving as an intricate system for coordinating and exchanging information at the microscale \cite{key5}. QS has the role of being a built-in system for networking artificial molecular communication candidates. The link between biological materials and an abstract mathematical concept such as signal processing and network coordination can give various new ideas for the development of more efficient signal processing systems for MC systems, including response thresholds and signal amplification.

Analyzing how these bacterial signaling molecules function in QS can help us understand the mechanism and could eventually lead to the creation of biochemical messaging protocols in MC systems. For developing nano-scale sensors and actuators, bacteria might be a good baseline with their ability to respond to environmental stimuli and features like specificity and sensitivity. Bacterial use of the signal molecules for communication and the collective building of the network models can help us study and develop decentralized communication models.

Hence, this section discusses the quorum sensing technique employed by bacteria and the specific signaling molecules involved in this mechanism. Some of the open issues are highlighted at the end of each subsection.

\subsection{Quorum Sensing}

Quorum sensing is a communication mechanism between cells that allows bacterial populations to achieve collective behaviors by sensing population density and bacterial composition changes \cite{key5}. It includes both processes of synthesis and identification of signaling molecules through their interaction with their environment \cite{key4}. Such molecules, when released by the bacteria, accumulate in the environment. Bacteria respond to these signals in a threshold-dependent manner. A molecule can be recognized as a QS signal if it accumulates in the extracellular environment, exceeds a certain threshold, and is sensed by a specific bacterial receptor \cite{key11}. This can happen in situations where the molecule is generated during a specific growth stage or as a response to environmental changes. When bacteria activate QS modes, they coordinate their behavior in response to the changing population density and species composition in their surroundings. In fact, not all cells in a bacterial community need to be in quorum sensing mode \cite{key11}. However, when fluid flow increases, the bacterial biomass required to elicit quorum sensing in the population increases. Gram-negative and Gram-positive bacteria use different QS mechanisms as depicted in Figure 2.

When bacteria activate their QS mode, they react in a coordinated manner and employ diverse signaling molecules \cite{key6}. In the MC context, the term ‘molecule’ may have a broader meaning than the word itself \cite{key96}. Regarding bacteria, ‘molecule’ might refer to the information-carrying molecules they secrete, such as autoinducers or cyclic diguanosine monophosphate (cyclic di-GMP). Alternatively, it can also refer to the bacterium itself, which serves as the carrier of information.

\begin{figure}[ht]
   \includegraphics[width=1\linewidth]{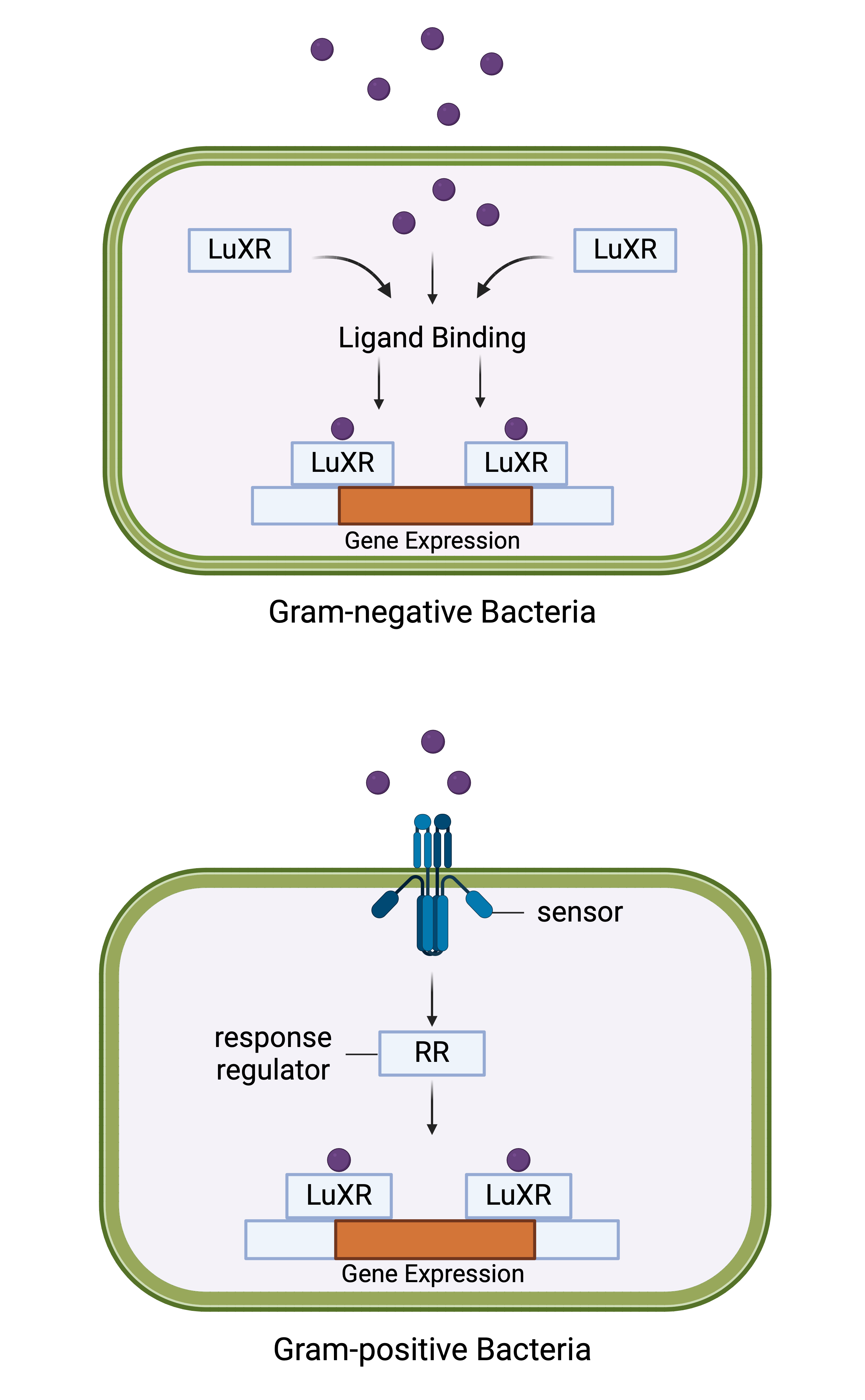}
    \caption{Comparison of quorum sensing methods in Gram-negative and Gram-positive bacteria. The top one illustrates the LuxR-type quorum sensing mechanism in Gram-negative bacteria. In this system, the LuxR protein attaches to autoinducer molecules that diffuse through the cell membrane. This binding process helps regulate gene expression. The bottom one illustrates a two-component system seen in Gram-positive bacteria. In this system, the sensor kinase recognizes autoinducer molecules and phosphorylates the response regulator (RR). The phosphorylated RR then binds to the LuxR protein, regulating the expression of genes. (Created with BioRender.com).}
\end{figure}

Bacteria use QS to communicate both with their own population and with eukaryotes such as plants \cite{key140} and mammalian cells \cite{key141}. Moreover, viruses are also able to detect and respond to certain QS molecules that influence their behavior \cite{key139}. The combined effects of signal molecule propagation distance and cellular responsiveness to these molecules determine communication range in QS \cite {key189}. The bacterial ability to recognize and react to signals from a variety of entities demonstrates the profound and adaptable nature of bacterial communication systems, underscoring their importance in the wider realm of MC and interactions between different species.

Quorum sensing is the switch from individual to collective behaviors via the number of cells in any given environment, and this approach provides insights into the design of communication networks that are both scalable and variable. In fact, bacteria are able to use QS to synchronize their behavior and adapt to their environment. Dynamic adaptations and self-organizing networks place bacteria as next-generation tools for communication systems that have to withstand ever-changing environments.

Although quorum sensing is one of the most studied topics in bacterial communication, current bacterial network models rely on oversimplified assumptions. Therefore, further in-depth studies and experiments are required as outlined below.
\begin{itemize}
    \item \textbf{Communication protocols:} In order to enhance the network efficiency, self-organization, and adaptability of the bacterial communication networks, communication protocols are required that mimics the QS mechanism. 
    \item \textbf{Signal propagation and reception modeling:} Accurate models are needed to see the signal propagation and reception in QS mechanism. Although there are models that shows diffusion of one signaling molecule, bacteria receives several signaling molecules simultaneously. Modeling of such networks is required to understand and represent bacterial communication precisely.
    \item \textbf{Energy-efficient communication systems:} Taking advantage of the bacteria's energy-saving strategies in QS might be critical, especially for some IoT applications. Therefore, it is needed to design networks that can mimic this mechanism in order to develop low-energy communication systems.
    \item \textbf{Microfluidic systems representing QS:} In order to represent the QS mechanism closer to the real scenario, it is necessary to create microfluidic systems. In this way, the designed models can be experimentally validated and improved. In addition, different diagnostic tools can be created using these systems.
    \item \textbf{Cross-species MC systems:} To develop cross-species MC systems, it is necessary to research and model the signaling pathways of various other organisms, such as plants and fungi, that QS molecules affect. Potential insights that could be gained from this could be useful for agricultural applications.
\end{itemize}

\subsection{Signaling Molecules}

Bacteria are classified into two groups, Gram-negative and Gram-positive, according to the structure and content of their cell walls. The cell wall of Gram-negative bacteria is very thin, which begins with peptidoglycan and continues with an outer membrane containing lipopolysaccharide \cite{key7}. In contrast to Gram-negative bacteria, Gram-positive bacteria have a thick cell wall composed of peptidoglycan substances \cite{key7}. This difference in their cell wall structure affects their reaction to antibiotics and adaptation to their surroundings. Moreover, different types of signaling molecules might be used in QS, depending on this distinction. For instance, Gram-negative bacteria utilize acyl-homoserine lactone (AHL) signaling molecules, while Gram-positive bacteria employ the peptide signaling mechanism in QS \cite{key8}.

Gram-negative bacteria uses autoinducers (AIs) as signaling molecules. The concentration of these molecules in the environment increases as the density of cell population increases \cite{key9}. There are two kinds of receptors that detect these AIs which are cytoplasmic transcription factors or transmembrane two-component histidine sensor kinases \cite{key11}. In contrast, Gram-positive bacteria use oligopeptides as AIs and its receptor is a two-component histidine sensor kinases \cite{key11}. Although Gram-negative and Gram-positive bacteria differ in many aspects, they both show common ground in one quorum sensing molecule which is AI-2 \cite{key60}. In other words, AI-2 serves as a universal language, allowing multi-species communication regardless of their Gram classification. Besides AIs, bacteria can synthesize second messenger molecules like cyclic di-GMP and use them to regulate different cellular functions such as motility, pathogenicity, and biofilm formation \cite{key10}. These molecules and their signaling pathways has been extensively studied as can be found in \cite{key9,key8,key10,key11}. In contrast to these studies, this study concerns with a focused approach to evaluate these signaling molecules in a particular context of molecular communication.

\begin{table*}
\caption{Bacterial Signaling Molecules Space}
\renewcommand{\arraystretch}{1.5}
\resizebox{\textwidth}{!}{%
\begin{tabular}{|m{1.4cm}|c|m{6cm}|m{2cm}|}
\hline
\textbf{Signaling Molecule Type}               & \textbf{Examples}                                                                                                                         &
\centering \textbf{Functions}       & \textbf{References}                                                                           \\ \hline
\textit{\textbf{Autoinducers}}    & \begin{tabular}[c]{@{}c@{}}AHLs (N-acyl homoserine lactones),  \\ AI-2\end{tabular} & \begin{tabular}[c]{@{}c@{}}Facilitate intercellular communication and control \\ the expression of genes\end{tabular} & \begin{tabular}[c]{@{}c@{}} \cite{key8}, \cite{key9}, \cite{key11}, \\ \cite{key60}, \cite{key149}, \cite{key150} \end{tabular}\\ \hline
\textit{\textbf{Second Messenger Molecules}}           & Cyclic di-GMP                                                                                                & \begin{tabular}[c]{@{}c@{}}Regulation of biofilm formation, motility, virulence, \\ cell cycle and differentiation\end{tabular} & \begin{tabular}[c]{@{}c@{}} \cite{key10}, \cite{key151}, \cite{key152} \end{tabular}                   \\ \hline
\textit{\textbf{Peptide Signaling Molecules}}  & \begin{tabular}[c]{@{}c@{}} Oligopeptides, Nisin, Subtilin \end{tabular}                                       & \begin{tabular}[c]{@{}c@{}}Antibiotic activity, regulation of gene expression \\ related to sporulation and competence\end{tabular} &  \begin{tabular}[c]{@{}c@{}} \cite{key4}, \cite{key11}, \cite{key153}, \\ \cite{key154}, \cite{key155}    \end{tabular}        \\ \hline
\textit{\textbf{Pheromones}}         & \begin{tabular}[c]{@{}c@{}}Competence and \\ sporulation factors \end{tabular}                                 & \begin{tabular}[c]{@{}c@{}}Luminescence, virulence factor production, fruiting \\ body development, competence, and sporulation\end{tabular}    & \begin{tabular}[c]{@{}c@{}}\cite{key156}, \cite{key157}, \cite{key158}       \end{tabular}                  \\ \hline
\textit{\textbf{Indole-based Molecules}}& \begin{tabular}[c]{@{}c@{}}Indole, Indole-3-acetic acid \\ 4-Fluoro-indole, Desformylflustrabromine \end{tabular}                                 & \begin{tabular}[c]{@{}c@{}}Regulates behaviors like pathogenesis, eukaryotic\\ immunity, antibiotic resistance, and biofilm formation \end{tabular}    & \begin{tabular}[c]{@{}c@{}}\cite{key221}, \cite{key222}, \cite{key225}, \\ \cite{key226}, \cite{key227}    \end{tabular} \\ \hline
\textit{\textbf{Volatiles}}& \begin{tabular}[c]{@{}c@{}}2,3-Butanediol, Acetoin, Histamine,\\ Dimethyl disulfide/trisulfide, Acetic acid \end{tabular}                                 & \begin{tabular}[c]{@{}c@{}}Inter-domain and inter-species communication,\\ timing of biofilm formation and defense mechanisms \end{tabular}    & \begin{tabular}[c]{@{}c@{}}\cite{key223}, \cite{key224}, \cite{key228}, \\ \cite{key229}, \cite{key230}, \cite{key231}    \end{tabular} \\ \hline
                                 
\end{tabular}%
}
\end{table*}

Bacteria use various types of receptors for signal detection. Some bacteria may use similar signals for QS, but the varieties of receptors that receive these signals and initiate subsequent signal transduction pathways are different \cite{key12}, so they function as multi-sensors. Researchers have discovered that different types of bacteria can recognize different signaling molecules at slightly different cell densities. As a result, different types of bacteria will react differently to various combinations of signaling molecules \cite{key13}. Bacteria have the ability to distinguish between signals coming from members of their own species (kin) and signals from members outside their own species (nonkin) \cite{key13}. Recognition of kin allows the formation of bacterial groups with genetically related members in which collaborative functions outstrip the abilities of an individual bacterium \cite{key14}. Furthermore, the changes in the numbers of both the kin and nonkin bacteria need to be carefully and simultaneously considered in order to equally distribute the public goods and prevent the exploitation by other organisms \cite{key15}. This objective can be implemented by maintaining a fine balance between sensitivity (the ability to detect signals with low concentrations) and specificity (the ability to discriminate between different signals). It is noteworthy that some bacterial species can enable a unidirectional communication process, which means some of them either show incapacity of AIs production along with the ability to detect it, or they produce AIs but have no detectable autoinducer receptors for reception.

Apart from these QS and second messenger molecules, there are various signaling molecules used by bacteria. Bacteria can also secrete these signalling molecules as a response to external influences such as stress and the presence of other microorganisms.  Indole and its derivatives plays a critical role in bacterial pathogenesis, eukaryotic immunity and communication with other organisms such as insects and plants \cite{key221, key222}. Moreover, volatiles, which are signaling molecules that serve as a potential common language for bacteria to communicate with other organisms such as animals, plants, and fungi, are involved in relatively long-range communication due to their highly diffusable nature \cite{key223, key224}. A summary of the signalling molecules mentioned in this article can be found in Table 1. As mentioned before, bacteria are involved in complex interactions with the variety of signalling molecules they utilize, which highlights its sophisticated nature. Aside from the identified compounds, there may be more that remain undiscovered.  

Further research is still required to gain a deeper understanding of these signaling molecules and their potential use cases. Some open directions are listed below.
\begin{itemize}
    \item \textbf{Impact of environmental factors:} In order to develop more robust MC models using bacteria, it is necessary to investigate the effect of environmental variables such as pH, other microorganisms in the environment, and exposure to chemicals on signaling molecules.
    \item \textbf{Decoding bacterial language:} By using semantic communication, which targets the meaning and context of the messages in bacterial signaling, we might end up getting to a much deeper level of understanding of the role of specific signaling molecules in bacteria or in their interaction with other organisms. This could lead to a better understanding of the behavior and reactions of bacteria.
    \item \textbf{Multimodal models:} Bacteria receives more than one input in their natural habitat. In order to analyze and mimic its behavior to these inputs, multimodal signaling models should be designed.
    \item \textbf{Testbeds:} Testbeds are crucial since they allow for reproducible and accurate experiments on systems that use biological entities. Various testbeds must be designed to study the dynamics and processes of signaling molecules and to improve our understanding of bacterial communication. In this way, targeted antimicrobial therapies, biosensors that respond to environmental cues to detect contaminants, and novel bio-computing systems can be designed and experimentally validated.
\end{itemize}

\section{Signal Transduction Mechanisms}

Bacterial signal transduction mechanisms are a complex network of processes that enable these organisms to sense and respond to environmental changes. This series of molecular processes includes signal reception by the receptor and transmittance of the signal through a series of biochemical reactions followed by a specific cellular response. This process is pivotal for the survival and adaptation of bacteria as well as getting used as an important piece of information in the area of MC and the generation of sophisticated methods of signal processing and transmission systems.

Bacteria are able to communicate with each other or orient themselves based on gradients. Moreover, they can interact electrically by using ion flow or potential changes across the cellular membrane by forming biofilms. Covering these bacterial mechanisms is of great importance for autonomous agents to move through complex environments and for fast MC systems based on electrical impulses. In this section, the signal transduction mechanisms that bacteria use are discussed.

\subsection{Bacterial Communications via Chemotaxis}

The habitat of bacteria is very complex and diverse. They possess two features that are crucial for survival, i.e., perception and interpretation of the environment, closely followed by an appropriate response. Chemotaxis refers to the movement of bacteria from higher concentrations of beneficial chemicals to lower concentrations of harmful substances \cite{key16}. The bacteria alternate between turns and run to move by means of their chemoreceptors to sense concentration gradients. This approach implements a temporal sensing technique to trace the flux in spatiotemporal concentrations \cite{key17}. In order to decide their direction, bacteria simply have to sense their immediate environment and foresee the near future.

The process of comparing and deciding accordingly leads to a behavior called ‘run and tumble’ \cite{key18}. In the “run” phase, bacteria show sustained forward motion, whereas the “tumble” phase is characterized by occasional shifts in orientations (Figure 3). If the concentration increases in the direction of travel, they persist in their ‘run’. They have a tendency to change direction when the concentration increases or remains constant, thus producing a ‘tumble’. Their flagella power the movement process \cite{key104}. The flagella rotate in the counterclockwise direction to produce thrust and keep the swimming direction in a straight line. When flagella rotates counterclockwise, it produces drag, and thus the direction of the movement changes \cite{key18}. The motility of the bacteria has been thoroughly studied in \textit{E. coli} \cite{key19}, revealing the role of membrane receptors and internal proteins, such as CheA, CheW, CheZ, and CheY, in modulating flagellum motor activity and consequently the movement of the bacteria. Moreover, while single cells display swimming motions in a liquid environment, groups of cells exhibit swarming motions \cite{key20}.

\begin{figure}[ht]
   \includegraphics[width=1\linewidth]{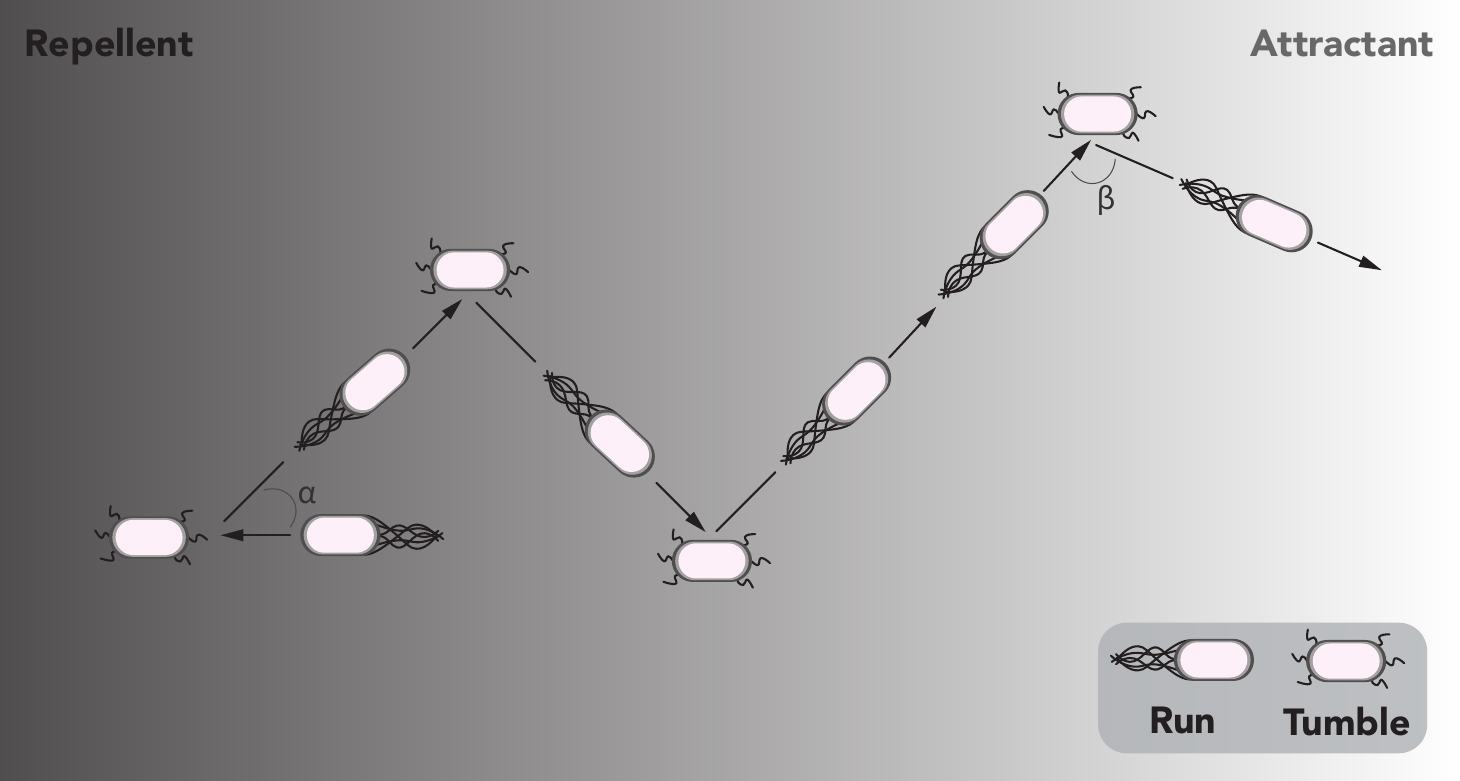}
  \caption{Run and tumble movement of bacteria in chemotaxis \cite{key104}. Change of flagella according to "run" or "tumble" movement is shown in the legend. Cells tend to keep moving forward as they rush toward attractants; when swimming away, they tend to flip over and change direction in response to repellents. The angle at which a bacterium changes direction during the "tumble" phase depends on various factors \cite{key80, key122, key233}.}
\end{figure}

The chemotaxis models frequently used in the literature usually assume that cells are able to detect concentration differences accurately \cite{key21,key22}. It is also assumed that the signaling molecules are constantly present in the environment at a detectable level. It must be noted that these chemical gradients are often very diluted \cite{key23}, which explains why they are so difficult to detect in the natural habitats of wild bacteria in marine and soil environments. Under such conditions, the bacteria are exposed to a signal that is noisy and scattered.

Noise, whether originating internally or externally, significantly impacts the range and accuracy of gradient sensing \cite{key182,key183}. Internally, fluctuations of signals take place within the cell and at the interfaces with receptors that receive signals. Externally, the noise results from other molecules in the external environment, especially at places where chemoattractants are present. A major strategy that bacteria employ to improve their chemotaxis efficiency is the successful management of noise. Bacterial communication often takes place in a stochastic environment where signals are weak and prone to noise. It is in these conditions that there is intentionally the occurrence of noise in receptor-ligand binding \cite{key136} resulting from known inherent stochastic variations and binding instability \cite{key123}. While the specificity of binding improves when bacterial receptors are engineered through synthetic biology, expecting these receptors to function optimally in dynamic environments may not be ideal. The state-of-the-art stochastic resonance behavior of bacteria provides a basis for developing more sophisticated signal amplification algorithms.

Further research is needed to explore bacterial chemotaxis in the following directions.
\begin{itemize}
    \item \textbf{Realistic models:} Although current models provide beneficial insights about the behavior of bacteria, most of them oversimplify the bacterial processes. More realistic models need to be created for applying and testing these models in practical situations. So, the investigation of this area must be continuous in order to develop a deeper understanding of bacterial intricate surroundings.
    \item \textbf{Synthetic chemotaxis system:} It is necessary to develop artificial bacterial systems in order to investigate and control the chemotaxis response. This may include ways to synthesize the artificial gradients that allow us to explore the principles of chemotaxis and use them in biotechnology and medicine. Moreover, studying the single-cell dynamic of the bacterium can reveal the decision-making mechanism of bacteria.    
    \item \textbf{Bacterial memory:} Bacteria's ability to recall the recent past can be used for tasks such as soil health monitoring in agriculture or remembering and reacting to specific conditions such as moisture levels or the presence of pests. As a result, increased resource efficiency and improved crop management can be achieved. Other than agriculture, it also serves as a method for smart drug delivery with the ability to remember certain conditions including pH level or biomarkers.
\end{itemize}


\subsection{Bacterial Communications via Electrochemical Signals}

Despite being classified as single-cell organisms, bacteria commonly engage in social interactions in their natural habitats \cite{key31}. Biofilms are highly organized structures of bacteria that provide protection for them with a self-produced matrix called extracellular polymeric substance (EPS) \cite{key36}. The similarities between biofilms and multicellular organisms have been studied to find a resemblance between them, given their intricate physiological and structural complexity.  

Bacteria gain important advantages when they move from their planktonic state to their biofilm-forming state. Within the biofilm, bacteria develop a higher tolerance to several instances, such as food scarcity, pH changes, and antibiotics \cite{key36}. The allocation of tasks within the biofilm enables bacteria in specific locations to undertake particular responsibilities. The interior of the biofilm provides an ideal setting for facilitating the transfer of genetic material between bacteria \cite{key36}. This increased level of interaction enables the transfer of information, such as antibiotic resistance and insights about the environment. Consequently, the organisms living in the biofilm develop into a strong and unified whole, aware of their surroundings. This cooperation, although beneficial for bacteria and systems exploiting them, comes in danger when considering the context of diseases \cite{key35}.

\begin{figure}[ht]
   \includegraphics[width=1\linewidth]{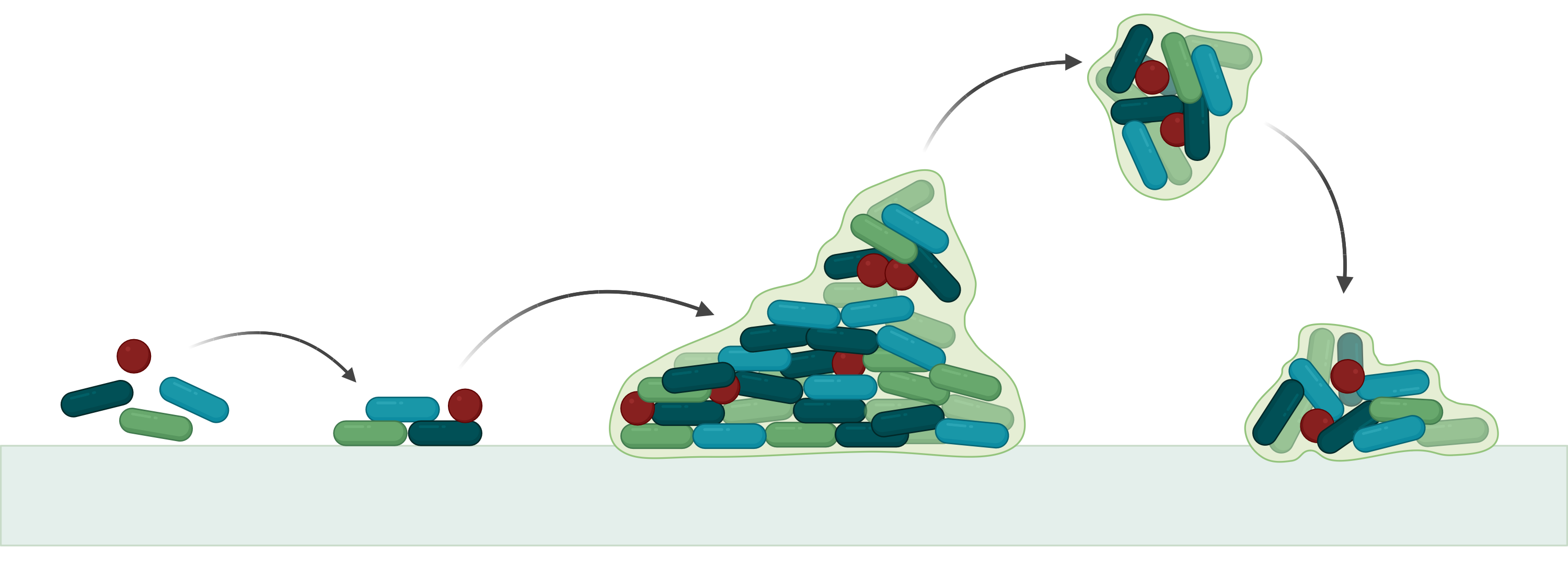}
    \caption{Illustration depicting the process of biofilm formation and growth. The graphic shows the first adhesion of planktonic bacteria to a surface, subsequent aggregation, and development into a sophisticated, multi-layered biofilm architecture. Following that, detachment events may take place, liberating individual bacteria and tiny clusters into the surroundings to spread and perhaps trigger new biofilm formation in other locations \cite{key235}.}
\end{figure}

The process of biofilm formation starts when a bacterium in a free-floating state comes into contact with a surface randomly or by directed motion (taxi) \cite{key32}. Some bacteria may utilize their pili as hooks, which enable the attachment process \cite{key33}. The initial attachment of bacteria may be reversible and depends on environmental conditions. Furthermore, higher adhesion stability comes from using the adhesin molecules \cite{key34}. Next, the bacteria get attached to the surface, and then they undergo a phase of division and growth, which is the definite structure that biofilms have. A simple illustration of biofilms is shown in Figure 4. The bacteria are known to communicate with each other not only by chemical signals but also through electrical signals in and out of the biofilms. Researchers have explored the significance of biofilms in promoting communication, in addition to their benefits and advantages for bacteria.

Nanotechnology techniques can be used to control and track bacterial biofilms in various biomedical processes. It is known that bacteria can be engineered and used for specific purposes \cite{key203}. By engineering the bacteria within biofilms with these techniques, it is possible to exert control over the characteristics of the biofilms. For instance, biofilms can be utilized to create novel and living composites \cite{key202}. These compounds are very innovative for biomedical applications, along with the theranostic ones, due to their capability to be used as dynamic responders, adaptability to the environment, and capacity for self-healing.

Ion channels are integral membrane proteins that function as the gateway for ions to pass through the membrane. They play a crucial role in promoting communication through electrical signaling, as shown in well-studied phenomena such as action potentials in neurons \cite{key97}. Bacteria possessing ion channels is an indication that bacteria are capable of more complex cellular mechanisms than portrayed in the past and can provide routes of communication. Empirical data indicate that bacteria can increase their sensing and control abilities by utilizing ion channels. Three target ion channels are used for mechanosensation in the \textit{E. coli} \cite{key29}. While these channels are usually interconnected with osmoregulation, cell wall production, and growth, experimental data shows their involvement in communication \cite{key38}. Studies have shown that bacteria have potassium, sodium, and chloride channels, along with calcium-gated potassium channels, and ionotropic glutamate receptors similar to neurons \cite{key30}. Bacteria within biofilms can establish electrochemical communication through numerous mechanisms \cite{key42}.

\subsubsection{Short-range Electrochemical Communication}

Bacteria are known for their capability of short-range communication \cite{key189} through diffusion, a phenomenon that has been previously modeled in the literature \cite{key41}. Several studies indicated that bacterial information carriers can perform intermediate-range communication through chemotaxis movement \cite{key39,key40}.

For their electrical communication over short distances, they use cytochromes \cite{key44} or bacterial nanowires. Even though diffusible molecules can be employed for short-range communication, electron transfer can have benefits over diffusible signaling, which can save time and energy without the risk of interference. Some \textit{Geobacter} species have demonstrated a direct electron transport ability to solid electron donors, acceptors, and even other cells \cite{key43}. The electrically conductive pili (e-pili) and c-type cytochromes allow direct electron transfer to occur. Among these, e-pili serve the key function of facilitating relatively long-range direct interactions. In addition, inner membrane cytochromes help electrons flow between the intracellular electron transport chain and periplasmic cytochromes, whereas periplasmic cytochromes allow the transfer of electrons between the inner and outer membranes \cite{key43}.

Bacteria form elongated, conductive protein filaments called nanowires that are able to transfer electrons from cell to cell over relatively large distances. These nanowires can be regarded as miniature electrical cables, which allow communication between bacteria. Early studies, using low-resolution imaging techniques, identified extracellular \textit{Geobacter} nanowires as type-4 pili (T4P) \cite{key45}. There are a number of research papers that investigate the structure, composition, and electron transport features of bacterial nanowires \cite{key46,key47,key48}, focusing on particular facets of the topic. The role of PilA in long-distance electron transport in \textit{Geobacter} is widely accepted. However, various studies show that under normal conditions, PilA is involved in the release of nanowires \cite{key49} rather than directly forming the extracellular conductive filaments \cite{key50}. The nanowires present on the surface of electricity-generating \textit{Geobacter sulfurreducens} are made up of cytochromes OmcS and OmcZ \cite{key51}. OmcS nanowires are essential for the development of electricity-generating biofilms during the initial growth stages, as well as in processes such as iron reduction and direct interspecies electron transfer \cite{key51}. In contrast, OmcZ nanowires are the key for the development of well-formed, mature biofilms with high current density \cite{key51}.

In the absence of direct contact, bacteria are capable of forming redox-active electron shuttles using soluble redox-active molecules \cite{key53}. Bacteria have the capability to produce and perceive these soluble redox-active molecules, which can undergo reversible electron transfer processes. In the case of \textit{S. oneidensis}, the redox shuttles can perform electron movement from bacteria to iron minerals \cite{key52}. \textit{S. oneidensis} is being researched in the field of biotechnology because of the exceptional capacity of the cells to transport electrons. This type of bacteria has the most significance in environmental contexts for all processes of metal and biogeochemical cycling of elements. Many uses are assigned to this characteristic, such as microbial fuel cells \cite{key98} and cleanup of polluted environments \cite{key99}.

\begin{figure}[ht]
\includegraphics[width=1\linewidth]{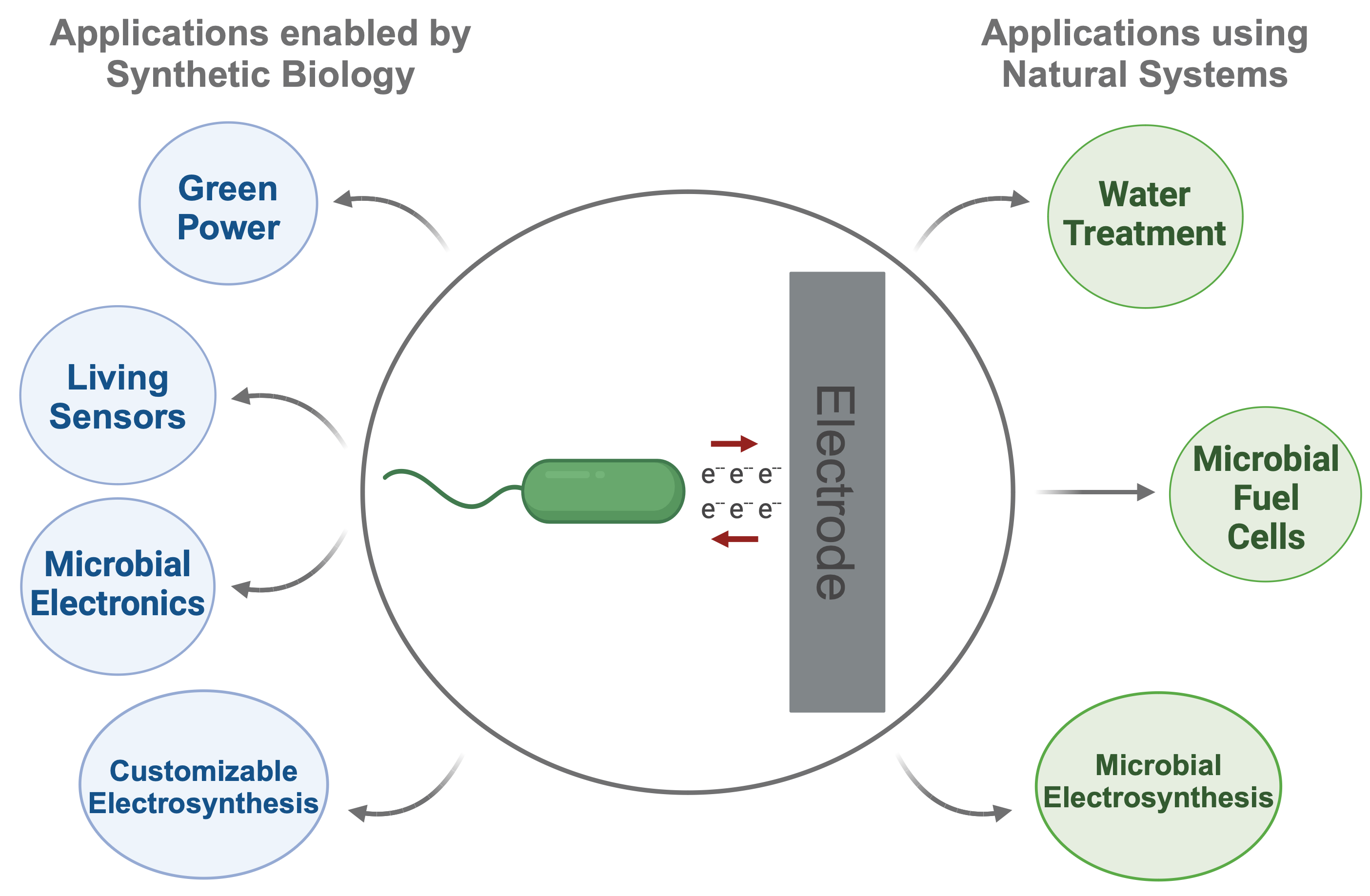}
    \caption{Current applications using naturally occurring electroactive bacteria and potential applications facilitated by synthetic biology \cite{key201}. Electroactive bacteria that occur naturally can serve as live catalysts for power production, water treatment, and microbial electrosynthesis. The application of synthetic biology tools enables the enhancement of power generation, the regulation and customization of microbial electrosynthesis, and the development of novel technologies such as sensors and living electronics. (Created with BioRender.com)}
\end{figure}

These electroactive bacteria, which include those found in the human gut \cite{key197}, have been the starting point of a number of emerging new research fields as shown in Figure 5 \cite{key200}. Genetic engineering is becoming the basis for future studies on the mechanism of electron shuttling and the creation of modified electroactive bacteria by means of optimization \cite{key198}. Several investigations have been conducted to enhance the connection between electrodes and electroactive bacteria in settings where they can be used \cite{key199}. Some of the studies focus on improving the electron flow in bacteria, particularly in manipulated strains like \textit{E. coli} that do not have the ability to transfer electrons naturally. The goal is to enable these bacteria to directly transfer electrons or to enhance the contact between the bacteria and the electrode. In order to enhance the transfer of electrons at the interface, it is necessary to develop novel materials that possess advantageous electronic properties.

\subsubsection{Long-range Electrochemical Communication}

Apart from other characteristics, one of the most important features of biofilms is their role in long-range communication \cite{key54}. It has been observed that ion channels in biofilms may promote long-distance communication within the microbial communities \cite{key42}. \textit{Bacillus subtilis} biofilms using K$^+$ waves have been proven to be able to communicate with bacteria or other biofilms that are far away by exchanging signals \cite{key100}, beyond what can be achieved through diffusion or nanowires. Another study shows that bacterial cells can recognize and use the ion concentration gradient of K$^+$ and Na$^+$ produced by an electric field at the solid surface to get closer to the surface \cite{key193}. Moreover, independent motile cells of different species unattached to any biofilm can be drawn to biofilm via electrochemical signals \cite{key55}. This implication suggests that biofilms have the capacity to change not only their own species' behavior but also that of foreign bacteria. Significantly, this process does not depend on chemical reactions or the synthesis of molecules. Instead, it operates by exploiting a signal that is universally recognized by bacteria due to their membrane potential. Thus, bacteria have the capacity to react swiftly to the incoming signals.

One of the most important advantages of communicating through K$^+$ waves is speed. Bacterial communication, or MC in general, differs significantly from other forms of communication, such as wireless communication, by experiencing significant delays \cite{key195,key196}. In some of the bacteria-based models in MC, information molecules are transmitted from transmitter to receiver via diffusion \cite{key41,key83}. In these models, delays may occur due to the channel or during the reception process. The release, detection, and movement of molecules involved in MC can be slower compared to other methods, and these delays may persist for hours \cite{key82}. Therefore, attempting to establish reliable communication between two inherently unreliable entities is a challenge. In a typical bacteria-based model, a cluster of bacteria is being used instead of a single bacterium, as it lacks inherent reliability \cite{key194}.

In certain designed models, bacteria serve as information carriers \cite{key39, key77,key78,key79} instead of relying on the diffusion of signaling molecules. In such circumstances, the bacterium must get to the targeted receiver, or community of bacteria, to facilitate conjugation and introduce a delay in the process. As previously mentioned, bacteria propagate between points through various movements. The multi-hop communication systems have been designed \cite{key79}, taking into account the purpose of minimizing loss by limiting the distance the bacteria need to travel.

Previous research has demonstrated that wild-type bacteria increase their swimming speed in high concentrations to improve their chemotaxis performance. However, the accelerated swimming comes at a metabolic cost to the bacteria. It can be inferred that the higher the swimming speed, the more energy is needed for locomotion \cite{key24}. This behavior, known as chemokinesis,is studied in \cite{key25}. By taking advantage of the inherent features of wild bacteria strains, efforts have been directed toward external manipulation of the swimming speed of bacteria, such as \textit{E. coli}. The speed of \textit{E. coli}'s swimming was affected by transferring the marine bacteria’s proteorhodopsin gene to it, making \textit{E. coli} responsive to light \cite{key26}.

While the understanding of the underlying mechanisms behind these bacterial communication ways is still being explored, the methods that have been uncovered thus far are already adequate to create different research avenues. Some of them are listed below.
\begin{itemize}
    \item \textbf{Noise reduction with biofilms:} Electrochemical communication between bacterial biofilms is less affected by noise and interference than the diffusion of signaling molecules. Therefore, designing models that use biofilms as nodes in a communication network can minimize these problems. Another way of communicating between biofilms could be to create synthetic waveguides, potentially reducing the time required.
    \item \textbf{Biofilm-mediated electrochemical signals:} Just as eukaryotes in bacteria's environment, such as plants, are affected by the diffusible signaling molecules \cite{key190} of bacteria, it is likely that they are also affected by electrochemical signals, and vice versa. Thus, intra-kingdom communication may be possible using biofilms. In order to understand how they are affected by these signals, it is necessary to design both models and experimental studies. Moreover, bacteria and fungi have been shown to have a mutualistic relationship in their natural environment \cite{key191, key192}. With these biofilm signals, communication efficiency can be improved, or the communication between them can be manipulated as desired. 
    \item \textbf{Artificial intelligence aided models for nanowire networks:} Using artificial intelligence aided models to simulate and research the complex mechanisms of the bacterial nanowire network structure could pave the way for network design principles to be developed, leveraging the adaptive and self-organizing nature of these models.
    \item \textbf{Biofilm-based composites:} Biofilm-based composites may serve as a convenient alternative to wearable devices, which can be made in the size of a tattoo. These composites can be used to not only sense certain cues in the human body but also communicate and respond accordingly, providing useful information, e.g., the effectiveness of the drugs being administered. 
    \item \textbf{Optimization of signaling molecules:} It is necessary to use signaling molecule optimization to alleviate delays on the communication channel. Such optimization techniques include decreasing the size and weight of the molecule, increasing solubility, and thus providing fast transportation of the molecule across the medium.
    \item \textbf{Optimization of medium:} It was shown that a slight change in the polymeric solutions enwrapping the bacteria can affect their behavior \cite{key122}. Fluid properties could be altered in order to minimize the delay. Thus, the migration and diffusion of chemotactic cells need to be regulated by changing the content of the medium.    
    \item \textbf{Increasing swimming speed:} Engineering the genes of the bacteria that determine the movement of flagella or pili to achieve faster swimming speed may be a direction to minimize delays. Another approach could be using external stimuli such as light or electrochemical signals which also enhances the swimming speed.
\end{itemize}

\section{Information Transmission in Bacteria}

Bacteria can divide quickly, thus producing corresponding daughter cells in which genetic information is passed on vertically to the offspring. On the other hand, the bacteria can also share genetic information among themselves through the horizontal transfer of genes \cite{key107}. Primarily, there are three distinct types of horizontal transmission: conjugation, transformation, and transduction. DNA transfer between bacterial cells is facilitated by conjugation in which the bacteria share genetic material through direct cell contact, therefore constituting a mating system \cite{key106}. In transformation, the recipient bacterium takes up a single strand DNA and integrates it into the bacterial chromosome using homologous recombination \cite{key105}. In transduction, a bacteriophage infects the recipient bacterium and copies its DNA \cite{key106}. In addition to these mechanisms, there is evidence of bacterial gene transfer utilizing membrane vesicles \cite{key70}. 

These horizontal gene exchange processes are of much importance to them as they enable them to adapt and evolve for their survival \cite{key107}. This genetic exchange mechanisms are important in bacterial populations for ensuring the exchange of information among bacteria. This allows bacteria to pick up advantageous genes, for instance, antibiotic resistance or the ability to use particular nutrients \cite{key108}. Furthermore, these mechanisms are the examples of bacterial communication. Bacteria’s ability to share information extensively, whether between their own or different species, and the rather easy process of getting this done, marks bacteria as one of the most valuable elements in the construction of MC networks. Besides chromosomal DNA, bacteria possess a circular DNA known as plasmids \cite{key56}.The recombinant DNA approach is being used to seal the information molecule, genes in this particular application, into the plasmid \cite{key135}. Along with the plasmid's replication, the inserted gene gets duplicated which enables transmission of information to other bacteria \cite{key109}. This kinds of DNA is easy to manipulate because of its small size and extrachromosomal attribute \cite{key56}. Thus, plasmids serve as effective carriers of information within bacteria.

From a molecular communication perspective, bacteria exploit their naturally occurring genetic exchange mechanisms as advanced molecular communication medium and information exchange methods. Analysis of the naturally occurring process of data transfer mechanism can help us to attain a more accurate knowledge and possibly reproduce the protocols used by bacteria. In the literature, bacterial genetic exchange mechanisms, especially conjugation, play a prominent role in MC networks where bacteria are used \cite{key39, key40, key41}. This section will review conjugation, transduction, and genetic exchange through membrane vesicles which are considered to be important in terms of MC networks. Figure 6 represents these three mechanisms in the simplest forms.

\begin{figure*}[ht]
  \includegraphics[width=\linewidth]{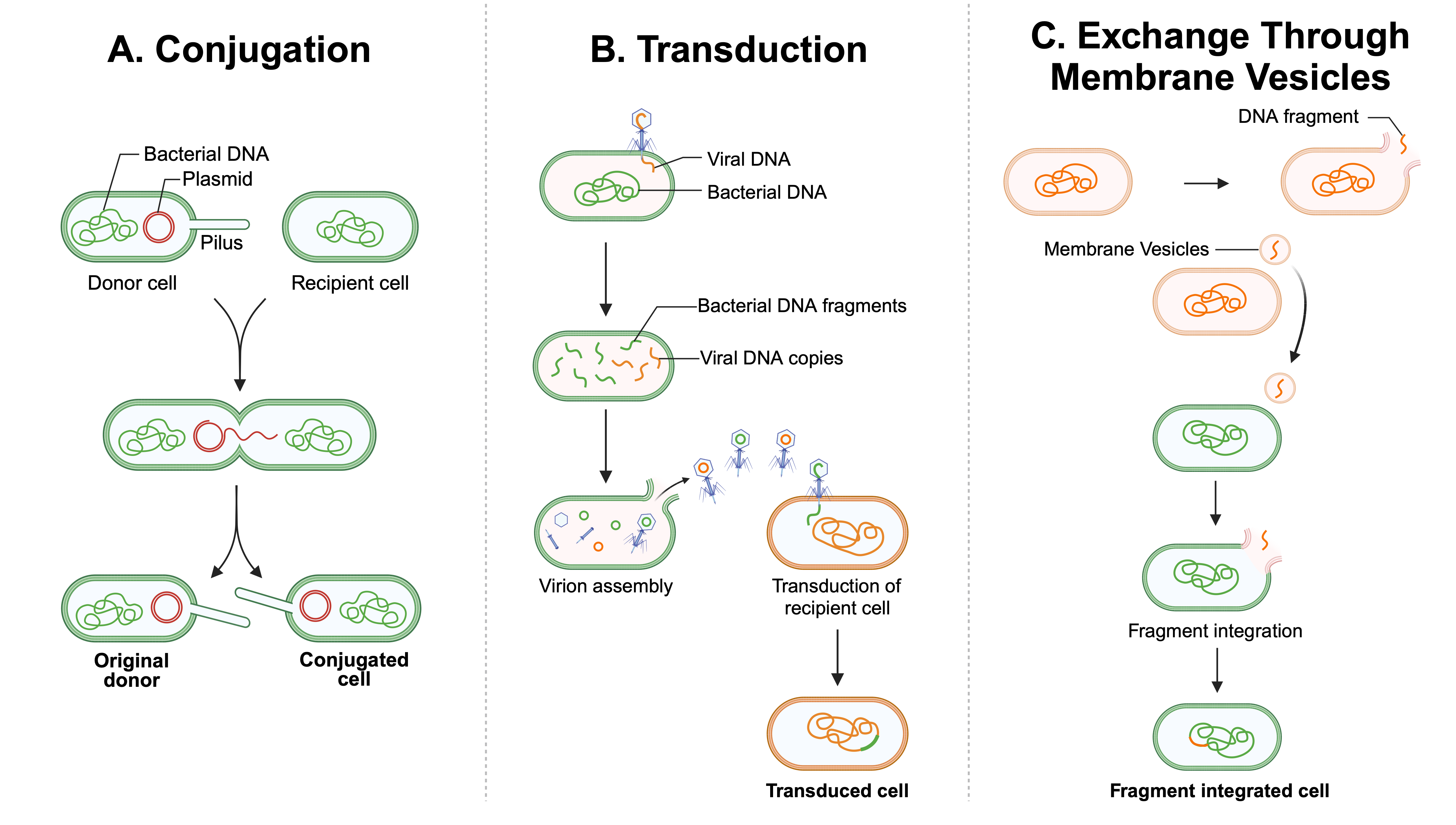}
    \caption{Genetic exchange mechanisms in bacteria \cite{key188}. (a) Conjugation. (b) Transduction. (c) Genetic exchange through membrane vesicles \cite{key236}.}
\end{figure*}

\subsection{Bacterial Conjugation}

Transfer of genetic information between bacterial cells via conjugation occurs through a one-way process that is enabled by direct contact \cite{key27}. The conjugation process is advantageous since it allows for information transfer between different species of bacteria. This transfer utilizes circular DNA units called plasmids that are able to replicate themselves independently from chromosomes. This conjugative plasmid contains the genes responsible for making contact with other bacteria and transferring their genetic material \cite{key28}. Transfer of genetic information happens via the use of the conjugation pilus, also known as the F pilus. A bacterium that has a pilus reaches out to another bacterium, where it latches on to it at the tip of the pilus and pulls it to create contact. Upon contact, the plasmid strand, which has been split by nuclease, is conveyed over to the other bacteria. As a result, both bacteria obtain the ability to assemble incomplete pieces of a complementary strand of plasmid with the use of the pilus gene that is transferred between the two bacteria. It is logical to assume that when some bacteria carry conjugative plasmids in a population and some do not, when a period of time passes, all bacteria in the community will have the conjugative plasmids.

Compared to other information transfer methods, conjugation stands out as a preferred mechanism since it is fairly direct and targeted. The cell-to-cell transfer observed in conjugation can be analogous to wired communication. It can be compared to a point-to-point communication link in engineering, where data is sent directly and securely between two nodes. The ability of the mechanism to be specific for certain biological aspects makes it uniquely accurate and selective in the passing of information. Nevertheless, it is still an inherent biological reaction and can carry potential drawbacks \cite{key184, key185}.

Conjugation rate is the frequency or speed of the transfer of information from a donor bacterium to a recipient bacterium \cite{key28}. Bacterial conjugation rate can be affected by different factors, such as the degree of taxonomic similarity in liquid matings \cite{key76}, which in turn has the potential to affect the efficiency of the network \cite{key186, key187}. It has been shown that conjugation between a bacterium and a population of bacteria is more efficient than conjugation between two individual bacterium \cite{key81}. This inference concerns not only the decoding of the messages but also putting a decoding delay into account during the process of information transmission when bacteria deliver the message or information to the designed receiver.

More studies are needed to both improve conjugation process and effectively use it as a way of reliable information transmission in molecular communication networks. Some open directions are listed below.

\begin{itemize}
    \item \textbf{Minimizing delay:}  When conjugation efficiency is low, it can introduce delays in the process, especially at the interfaces with artificial sensors. In order to minimize these delays, the conjugation rate could be enhanced by using bacterial biofilms. Whether by covering the sensor with a biofilm or conjugating the carrier bacteria with bacterial clusters, significant reductions in delays can be achieved. Moreover, biofilm provides a stable environment for the connection of pili.
    \item \textbf{Advanced genetic engineering:} It might be possible to monitor and manipulate the conjugation process through the use of advanced genetic engineering. Developing these techniques is needed to enable us to selectively activate and deactivate conjugation and regulate the transfer of specific information. 
    \item \textbf{Feedback mechanisms in conjugation:} Incorporating feedback mechanisms into bacterial systems to monitor and control the conjugation process may involve using sensory bacteria to assess the effectiveness of conjugation events and adjust future transfers accordingly.
\end{itemize}

\subsection{Bacterial Transduction}

Bacteria can transfer their genetic material by means of transduction. The process of transduction uses bacteriophages, also known as phages \cite{key69}. Bacteriophages are viruses that can replicate themselves within bacteria by hijacking their mechanisms. Some of the phages are able to promote the transfer of the DNA segment between different bacteria. Upon attachment, a phage introduces its nucleic acid into the host bacterium. Then, a phage enzyme cuts the DNA of the host cell into smaller pieces, which causes the bacteria's genome to unpack and the phage DNA to be replicated. As the DNA of the bacterium is also cut into smaller pieces, some phages encapsulate the DNA of the bacterium instead of their own and carry it to another bacterium. Thereafter, the phage goes and transfers this information to another bacterium, integrating the other bacterium's DNA piece into the receiver bacterium instead of its own DNA.

Transduction is a way of communication among bacterial populations that exploits a bacteria-to-phage link instead of bacteria-to-bacteria contact. While using transduction as an information transmission mechanism, it is crucial to have a well-designed system since the presence of phages might damage the bacteria. The function of bacteriophages that allow DNA conveyance and exchange affords the option to intervene and transmit information between populations of bacteria.

Although transduction is another way of information transmission between bacteria, more research is needed to use it as a reliable way of communication. Open directions for future investigation of bacterial transduction mechanisms include the following.
\begin{itemize}
    \item \textbf{Engineering the phages:} Taking up bacterial DNA by bacteriophages is a random process. In order to build a reliable communication system, bacteriophages can be engineered to pick up only a specific DNA piece from the bacteria. Moreover, engineering the phages according to our needs might also prevent them from causing harm to bacteria. Realizing this might enable us to use phages like data packets in communication.
    \item \textbf{Host selection:} Similar to other viruses, bacteriophages select their hosts in a species-specific manner. While this might be useful for information transmission throughout a single bacterial species, engineering the phages might allow for this transfer to be independent of the species of the bacteria. 
\end{itemize}

\subsection{Genetic Exchange Through Membrane Vesicles} 

In addition to the well-known genetic exchange mechanisms like conjugation and transduction, extracellular vesicles have also been proven to be a novel cell communication mechanism. Membrane vesicles are able to carry not only genetic information but also biologically active compounds such as proteins, lipids and carbohydrates \cite{key211}. These membrane vesicles acts like information cargo between bacteria \cite{key70} and involves the fusion of membrane vesicles that carries the genetic material of one bacterium to the recipient bacterium. Small membrane vesicles (MV), with sizes ranging from 20 nm to 300 nm, is produced at the growth phase of the bacteria \cite{key71}. 

Membrane vesicles can be taken up by several bacteria at once, since direct cell-to-cell contact is not required. Therefore, this method is analogous to broadcasting in wireless networks. Furthermore, this technology could be further elbowed through the use of biosynthesis to create specialized vesicles. In-depth understanding of the interaction and translocation of genetic material in the context of vesicles can be a basis for the usage of this mechanism in synthetic applications.

Although this is an established way of genetic exchange between bacteria, more research needs to be conducted in order to determine the feasibility of using this mechanism as a way of information transmission in communication networks. Some open directions are provided below. 
\begin{itemize}
    \item \textbf{Analyzing the transfer process:} It is known that membrane vesicles can easily be absorbed by bacteria, both Gram-negative and Gram-positive \cite{key210}. Using these vesicles might be one of the preferable ways of exchanging information, but there is an increasing demand for additional research to pinpoint the components that can influence the transfer process.
    \item \textbf{Cargo content:} Studies are needed that investigates the cargo transfer capabilities of chosen vesicles, so the content of the cargo could be altered and manipulated by modifying the vesicles. 
    \item \textbf{Computational models:} Computational models are needed to simulate the behavior of membrane vesicles in an MC network. This can help understand the interaction with target cells and optimize for effective communication.
    \item \textbf{Data networks with MVs:} MVs can be engineered to operate as nodes within a dynamic data network. Each vesicle may contain distinct genetic information that, when coupled with others, creates a sophisticated dataset or message.
\end{itemize}

\section{Bacterial Traits and Their Specific Roles in Molecular Communication}

Bacteria are the subject of many MC studies with their adaptability, intricate sensing mechanisms, and relatively complex communication skills. Their inherent biological signaling pathways and ability to regulate collective behaviors make them perfect candidates for MC networks. Nevertheless, bacteria do not constitute a homogeneous species; they comprise several species with varied properties. With their diverse appendages and apparatuses, they function as information carriers or even the information itself in bio-inspired communication systems. An in-depth understanding of these unique traits is needed to build communication systems according to one's needs. This section highlights the most important traits of bacteria for them to be utilized in MC systems. Existing challenges and possible future directions are investigated at the end of this section. 

\begin{table*}[ht]
\caption{Bacterial traits, their biological significance, and role in molecular communication frameworks}
\label{your-table-label}
\centering
\renewcommand{\arraystretch}{2}
\resizebox{\textwidth}{!}{
\begin{tabular}{|>{\bfseries\raggedright\arraybackslash}m{1.8cm}|>{\raggedright\arraybackslash}m{7cm}|>{\raggedright\arraybackslash}m{7cm}|}
\hline
\rowcolor{lighterheadercolor}
\textbf{Bacterial Traits} & \multicolumn{1}{>{\bfseries\raggedright\arraybackslash}m{7cm}|}{\cellcolor{lighterheadercolor}Biological Significance} & \multicolumn{1}{>{\bfseries\raggedright\arraybackslash}m{7cm}|}{\cellcolor{lighterheadercolor}Role in Bacterial Molecular Communication Networks} \\
\hline
Number of Plasmids & Coexistence of more than one plasmid within a bacterium may influence the efficiency and specificity of the conjugation \cite{key56} & Due to variations in resource allocation and the existence of specific transfer mechanisms encoded by each plasmid, having more than one plasmid can lead to interference or loss of information in MC networks \\
\hline
GC Content of the Plasmid & Relative abundance of guanine (G) and cytosine (C) nucleotides in the DNA strand can effect the efficiency of conjugation, plays a significant role in the stability and compatibility with the host cell \cite{key220} & Gene transfer between bacteria that have similar GC content in their plasmids can improve the conjugation efficiency, which in turn enhances the MC by increasing the likelihood of successful information transmission \\
\hline
Pilus & In order to transfer plasmids, bacteria needs to possess a pilus which are filamentous appendages that have functions such as motility, DNA uptake and biofilm formation \cite{key57} & Pilus can function as conductive wires in the electrical network analogy by allowing the transmission of information between different bacteria \\
\hline
Flagellum & Bacteria needs flagellum to perform run-and-tumble action as part of the chemotaxis, which is an elongated, whip-like appendages primarily used for motility \cite{key57} & In MC networks where bacteria are used as information carriers, flagellum is the appendage that can actuate movement by perceiving their surroundings and navigating in a direction that is beneficial \\
\hline
Non-competency & Some bacterial species have the ability to take up naked DNA that is not enclosed in a membrane and incorporate it into their genetic material through recombination process \cite{key101} & Non-competency needed in MC networks where it is not beneficial for bacteria to take up naked DNA, since it may cause interference \\
\hline
Surface Exclusion & Using a surface protein, bacteria has the capacity to hinder the transfer of external DNA, such as plasmids \cite{key62} & Surface exclusion is critical in networks where bacteria act as transmitters of information or where it is undesirable for bacteria to exchange genes with one another in order to prevent noise and interference \\
\hline
\end{tabular}
}
\end{table*}

\subsubsection{Number of Plasmids}

In a communication network utilizing bacteria, plasmids operate as the message that needs to be transmitted to the receiver via various genetic exchange mechanisms. The characteristics of plasmids and their behaviors during transmission process is essential for both engineering purposes and monitoring the information flow. The coexistence of numerous plasmids within a bacterium can have an influence on both the efficiency and specificity of the communication system \cite{key56}. Moreover, the competition for resources and replication machinery amongst plasmids might hinder the transfer of certain plasmids or lead to their direct loss. These situations require thoughtful analysis, as they have the potential to result in the loss of information within the designed system.

\subsubsection{GC Content of the Plasmid}

The GC content is among the factors that could determine the efficiency of conjugation. It is the measure of the ratio of guanine (G) and cytosine (C) nucleotides in the bacterial DNA string \cite{key132}. The GC content of bacteria determines their stability and a number of physical characteristics. Moreover, GC content can influence the compatibility of the bacteria with the host cell. In terms of conjugation, which has been defined as the information transmission method in bacterial networks, higher compatibility with the host may increase the efficiency. While designing communication systems, choosing plasmids that have similar GC content can prevent possible delays that come from the conjugation process.

\subsubsection{Pilus}

Pili are one example of the filamentous appendages that have a diameter of 5-8 nm \cite{key207}, the roles of which in bacteria include conjugation \cite{key57} and beyond. The functions of pili include adhesion, motility, DNA uptake, and biofilm formation \cite{key57}. Apart from these functions, they also serve as conductive appandages, particularly in a subset of bacteria such as electroactive \textit{Geobacter} \cite{key43} and \textit{Shawanella} \cite{key61}. The existing types of pili are wide-ranging due to their structure and mechanism of formation. Perhaps most widespread is type IV (T4P) for both Gram-negative and Gram-positive bacteria \cite{key205}. Besides bacteria, bacteriophages offer the ability to employ pili as receptors for therapeutic purposes \cite{key206}.

In conjugation, bacteria use their pili to transfer the genetic material, i.e., the information intended to be conveyed to other bacteria. Not all bacteria contain pili; hence, not all of them are able to transfer their plasmids via conjugation. Therefore, when bacteria are to be used in communication systems, especially where information transmission is to be done through conjugation, it is important to select bacterial species that possess this apparatus. Bacteria carry the gene for pilus production in their plasmids. Thus, when a donor bacterium conjugates with a recipient bacterium, this gene can be transferred to the other bacterium along with the plasmid. However, in populations of bacterial species that do not have the ability to conjugate, the transfer of information should be done in the ways mentioned previously, and the specified communication challenges should be taken into account.

\subsubsection{Flagellum}

The flagellum is another important appendage of bacteria with its ability to perform the run-and-tumble action as part of chemotaxis. Flagella are elongated, whip-like appendages primarily used for motility \cite{key57}. Bacterial motility comprises a range of movements, such as swimming, swarming, gliding, twitching, and floating \cite{key58}. Bacteria possessing flagella have developed a mechanism that enables them to perceive their surroundings and navigate in a direction that is beneficial. Therefore, in contrast to bacteria that do not have flagella, they do not rely exclusively on Brownian motion to navigate their environment. The transmission of information from a transmitter to a receiver in communication systems utilizing bacteria is dependent on the flagella. Certain communication networks utilize nodes, and bacteria can be used as carriers of information in between. In such systems, the dependence of bacteria's movements on flagellar motors rather than random motion is crucial for reliable communication.

\subsubsection{Non-competency}

In addition to these appendages, there are various other features of bacteria that can be beneficial in communication, such as non-competency. Competency refers to a state in which the cell wall and cell membrane have increased permeability, allowing cells to take in DNA \cite{key101}. Bacteria that possess this feature have the capability to take up naked DNA from the environment and incorporate it into their genetic material through recombination \cite{key59}. During instances of cell death and bacterial lysis, the information within the bacterium's plasmid is lost. Especially in a densely populated community, occurrences of cell death significantly increase the possibility of having contact with these DNAs that are not enclosed in a membrane. Even though this ability is beneficial for bacteria to have more knowledge about their environment, it can cause interference in communication models where nodes are utilized. An example of such a model is illustrated in Figure 7 \cite{key138}. Therefore, the lack of competence becomes a vital characteristic to take into account while choosing the bacterial species when designing communication models.

\subsubsection{Surface Exclusion}

Bacteria have the capacity to utilize a technique called surface exclusion, which might also be crucial for communication models. This method allows the bacteria to hinder the transfer of external DNA, such as plasmids \cite{key62}. A bacterial surface protein inhibits the entry of unnecessary conjugative material into the cell. This characteristic of bacteria hinders replication in situations where bacteria act as transmitters of information by decreasing interference or where it is undesirable for bacteria to exchange genes with one another.

\begin{figure}[ht]
   \includegraphics[width=1\linewidth]{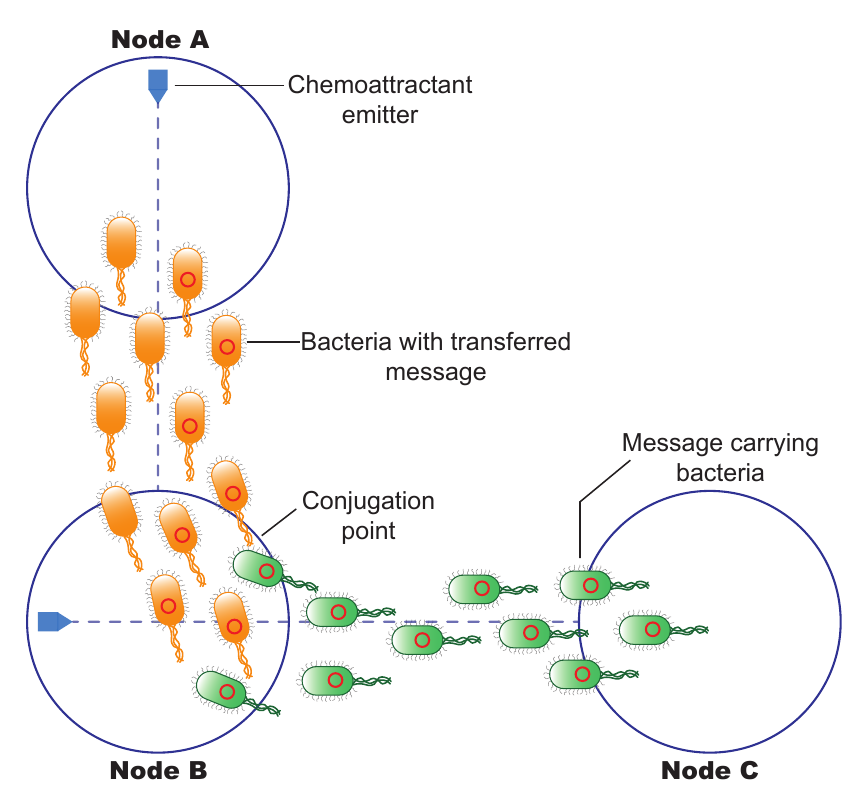}
    \caption{Bacteria acting as information carriers between different nodes \cite{key138}. Bacteria that carry messages are released from Node C and travel towards Node B, guided by chemoattractants. Upon reaching Node B, these bacteria transmit the message through conjugation. Subsequently, a different group of bacteria conveys this message from Node B to Node A, also directed by chemoattractants.}
\end{figure} 

In certain designed models, based on the intended application or hypothesis, some of the mentioned elements may not be utilized. Table 2 summarizes the bacterial traits mentioned in this section. However, in the majority of intricate bacterial systems, these characteristics are crucial factors to take into account. Some of the open directions for using these bacterial traits are listed below.

\begin{itemize}
    \item \textbf{Plasmid compatibility:} Plasmids that coexist or compete with each other needs to be identified, since it effects their stability and transmission.
    \item \textbf{GC content and genetic information transfer:} Further research is needed to understand how GC content affects gene expression and gene transfer. These aspects may influence bacterial communication and behavior.
    \item \textbf{GC content and bacterial adaptability:} Investigating the impact of GC concentration on bacterial adaptation can offer valuable insights into the strategies employed by bacteria.
    \item \textbf{Quantitative analysis of pilus:} There needs to be more methods available for the precise assessment of pilus function in bacteria. Advancing these methods would offer more profound understanding of pilus dynamics and their function in molecular communication.
    \item \textbf{Pilus in microbiota-host interactions:} Further research is needed to completely comprehend the prevalence, function, and importance of filaments in the high-density microbial communities of the gut \cite{key208, key209}. This will certainly result in new treatment approaches to adjust the composition of microbiota and improve the health of the host.
    \item \textbf{Pili as waveguides:} Since pili are conductive, they can function similarly to conductive wires in an electrical network analogy. This feature of bacteria raises questions about their potential to be used as waveguides. While they are not identical to standard waveguides, they can be engineered to function as such with their guiding properties. As opposed to traditional waveguides, these pili are susceptible to biological limitations, and their features can change based on a bacteria’s physiological status or external factors. Given the current state of research in this domain, it is imperative to thoroughly explore the underlying biological mechanisms in order to facilitate practical implementation.
    \item \textbf{Bacterial swarm intelligence:} Bacterial swarms can serve as a model for developing decentralized communication networks. Within these networks, individual agents function independently but can collaborate to accomplish intricate tasks, mirroring the self-organization and communication seen in bacterial colonies for resource distribution or mobility. These agents in the network might be powered by synthetic or bio-inspired flagellum. 
    \item \textbf{Flagellum for harsh settings:} The flagellum can be seen as a biological device that can actuate movement. Its capability to operate in harsh settings, such as at a pH of 2, could expand the range of potential applications for electrical systems created based on these biological structures.
\end{itemize}

\section{Bacterial Computing}

In nature, bacteria exhibit limited behavior, though more complex than initially perceived. Nevertheless, the progress in synthetic biology has demonstrated the ability to program bacteria, enabling them to perform specific, desired tasks. Parallel computation becomes necessary in problems where potential solutions grow exponentially with an increasing number of variables \cite{key127}. However, conventional computers are not always adequate for such processes. In contrast, living organisms function perfectly well in complex and multivariate environments and solve complicated problems in a reasonable amount of time. Therefore, biological agents, applicable in both nanoscale and macroscale networks, are proposed as an alternative computational paradigm \cite{key128,key129}.

Among them, bacteria stand out with their communication skills, intelligence, and problem-solving capabilities, leading to the field of bacterial computing. Indeed, bacteria possess the ability to function as a Turing-complete machine, enabling them to execute tasks that are comparable to those achievable by a modern-day computer \cite{key63}. So far, the use of bacteria has proven to be more efficient and faster than traditional computers in solving problems like the Hamiltonian \cite{key65} and the burnt pancake problem \cite{key64}. Bacterial species are able to adapt and evolve quickly, which makes them suitable to be used for solving many environmental problems \cite{key66,key67,key68}.

\begin{figure}[ht]
\includegraphics[width=1\linewidth]{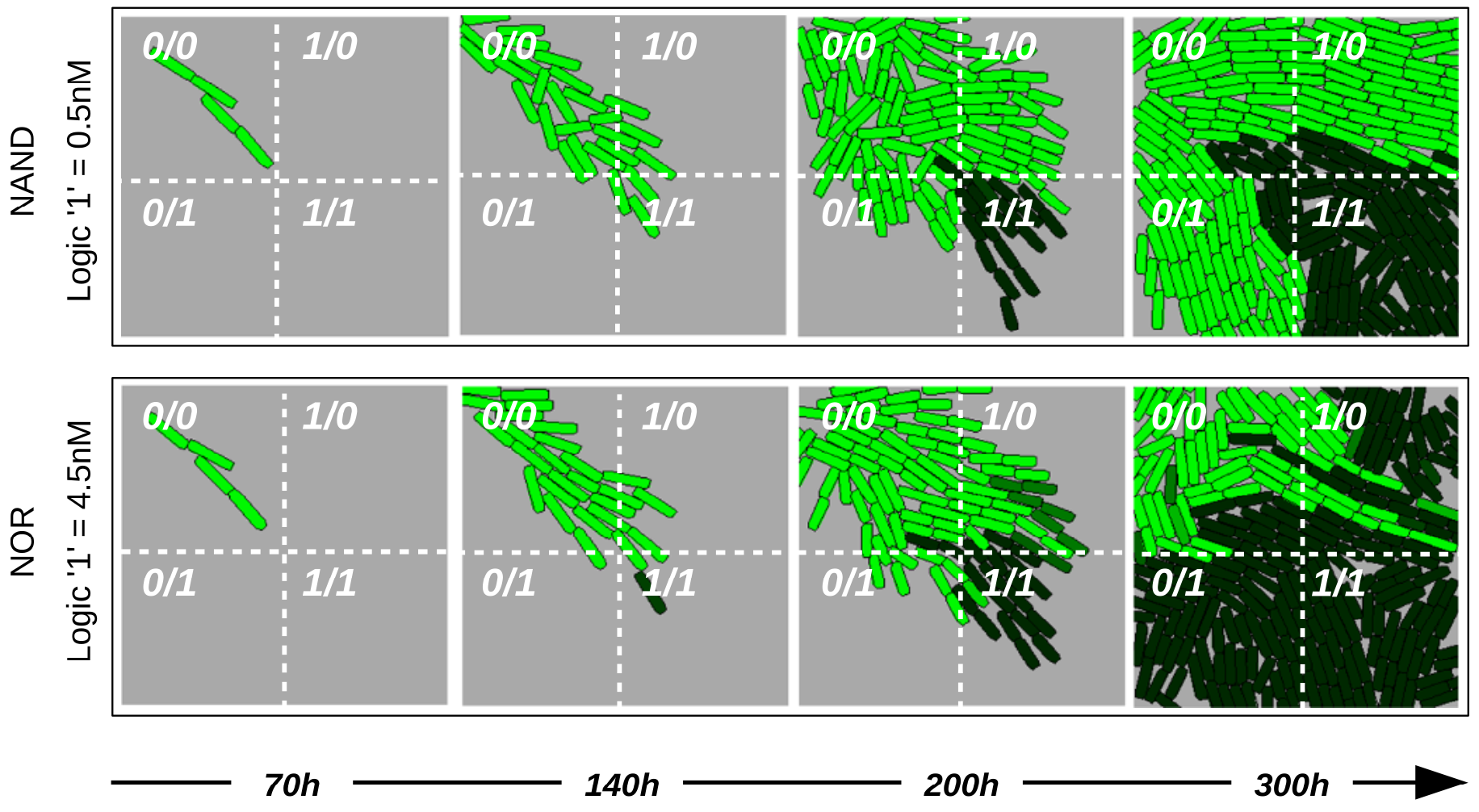}
    \caption{Simulation results of population based NAND and NOR gates using bacteria is shown. The columns represent different logic inputs (‘0’ means no input while ‘1’ means input of A or B is present) applied to the bacteria. Green fluorescent protein is used for visualisation purposes. For NAND, the bacteria grow in all conditions except when both inputs are 1 (1/1). For NOR, bacteria grow only when both inputs are 0 (0/0). This demonstrates how bacteria can be programmed to perform basic computations similarly to electronic logic gates. (Adapted from \cite{key234}).}
\end{figure}

With the advancements in synthetic biology, it is possible to artificially program, produce, standardize, and modify bacterial DNA pieces in order to create genetic circuits. By inserting this modified DNA into the bacterial cells, custom-built computers that work according to our needs are possible. So far, bacteria have been used to create logic gates (Figure 8) \cite{key74}, toggle switches \cite{key72}, oscillators \cite{key73}, and have shown the capability to store information and remember previous cellular events \cite{key75}. These are adaptations of genetic circuits to both digital and analog computing paradigms. It is worth mentioning that bacteria demonstrate these capacities not only in regulated surroundings but also under severe, toxic, and ever-changing conditions. Given the impracticality of operating electrical devices in such environments, there may be potential for utilizing systems constructed with bacteria in these conditions.

Our discussion in this section focuses on molecular machine learning mechanisms by comparing artificial neural networks with bacterial computing. Next, we examine several models that combine sensing and communication, emphasizing the collaborative usage of these two components in the context of bacteria. Finally, we also discuss quantum computing in bacteria, recognizing its significant potential. Current challenges and open directions can be found at the end of each section.

\subsection{Bacteria-Based Molecular Machine Learning}

The computational abilities observed in certain biological systems, such as bacteria, have given rise to the development of biological artificial intelligence \cite{key103}. Despite lacking a brain and not being considered conscious, bacteria exhibit minimal cognitive capabilities \cite{key102}, as evidenced by their ability to learn and adapt in various studies \cite{key133,key134}. Bacteria engage in molecular communication through both intra- and inter-cellular signaling. When these MC systems are engineered or integrated with artificial intelligence, more efficient and robust models emerge. There are various approaches to consider in this regard. One approach is to genetically engineer bacteria to create machine learning systems, while another approach is computing naturally occurring structures in bacterial cells. Studies involving artificial intelligence and machine learning (ML) focusing on bacteria have offered a new look into the field.

The idea that bacteria could have a network similar to artificial neural networks (ANNs) built from neuronal architectures \cite{key86} is becoming more plausible. Table 3 presents a comparison of similar principles found in ANNs and bacterial computing processes. Previous studies in the literature have made efforts to genetically modify bacteria in order to develop architectures similar to ANNs \cite{key85,key87,key88}. By leveraging the interactions and coordinated group behavior inherent in bacterial communities, perceptron networks have been developed. These perceptrons, composed primarily of four key elements – the input layer, weights and bias, net sum, and activation function – were engineered using bacteria. This innovative approach has proven effective in capturing the non-linear behaviors observed in biological systems. In \cite{key85}, a single-layer ANN utilizing \textit{E. coli} was constructed. This system employed a log-sigmoid activation function, wherein each cell received an inter-cellular diffusing molecule, triggering the expression of green fluorescent protein in response. In \cite{key87}, perceptron behavior was once again achieved by employing genetic circuit engineering in \textit{E. coli}. This demonstration illustrated the non-linear logarithmic input-output relationship within genetic circuits. In \cite{key102}, an offline-trained perceptron neural network was employed to program a bacterial population, simulated in silico. This marked a significant advancement in the utilization of genetic circuits to emulate perceptron behavior. In \cite{key88}, a consortia-based bacterial ANN was developed, experimentally tested, and distinguished by establishing a feedback mechanism between the receiver and sender cells through quorum sensing. These studies shows that the fundamental principles of ANN have been successfully applied to living bacteria.

\begin{table*}
\caption{Some of the common concepts of artificial neural networks and bacterial computing processes}
\renewcommand{\arraystretch}{1.5}
\resizebox{\textwidth}{!}{%
\begin{tabular}{|c|c|c|c|}
\hline
\textbf{Concepts}               & \textbf{ANNs}                                                                                                                              & \textbf{Bacterial Computing}                                                    & \textbf{References}                                      \\ \hline
\textbf{\textit{Parallel Processing}}    & \begin{tabular}[c]{@{}c@{}}Employs parallelism by having networked \\ neurons that process information simultaneously\end{tabular} & \begin{tabular}[c]{@{}c@{}}Parallel structure of bacterial colonies allows for \\ the simultaneous execution of several tasks\end{tabular} & \begin{tabular}[c]{@{}c@{}} \cite{key164}, \cite{key165}, \cite{key166}, \\ \cite{key167}, \cite{key168} \end{tabular} \\ \hline
\textbf{\textit{Adaptability}}           & Can adapt and learn from the data                                                                                                 & \begin{tabular}[c]{@{}c@{}}Can adapt by using genetic programming or \\ communicating with nearby bacteria\end{tabular}   & \begin{tabular}[c]{@{}c@{}} \cite{key159}, \cite{key160}, \cite{key161}, \\ \cite{key162}, \cite{key163}\end{tabular}          \\ \hline
\textbf{\textit{Distributed Computing}}  & \begin{tabular}[c]{@{}c@{}}Information is processed as a \\ distributed manner\end{tabular}                                       & \begin{tabular}[c]{@{}c@{}}Depends on bacterial colonies' distributed manner \\ for computation and communication\end{tabular}& \begin{tabular}[c]{@{}c@{}} \cite{key164}, \cite{key169}, \\ \cite{key170}, \cite{key171} \end{tabular}         \\ \hline
\textbf{\textit{Feedback Loops}}         & \begin{tabular}[c]{@{}c@{}}Use of feedback for learning \\ and adjusting the weights\end{tabular}                                 & \begin{tabular}[c]{@{}c@{}}Use of feedback for adaptation to \\ dynamically changing environments\end{tabular}   & \begin{tabular}[c]{@{}c@{}}  \cite{key172}, \cite{key173}, \cite{key181} \\ \cite{key174}, \cite{key175}      \end{tabular}                  \\ \hline
\textbf{\textit{Information Processing}} & \begin{tabular}[c]{@{}c@{}}Processing through weighted \\ connections and activation functions\end{tabular}                       & \begin{tabular}[c]{@{}c@{}}Processing through genetic circuits and\\  communication pathways\end{tabular} & \begin{tabular}[c]{@{}c@{}} \cite{key165}, \cite{key176}, \cite{key177}, \\ \cite{key178}, \cite{key179}, \cite{key180}    \end{tabular}                            \\ \hline
\end{tabular}%
}
\end{table*}

Beyond these studies, there are systems created by mimicking the structures naturally found in bacteria. Bacteria are able to identify their surrounding conditions including the concentration of other bacteria as well as changes in pH and temperature. This capability is due to their means of communication, which allows them to make well-informed judgments regarding the observed changes. Nevertheless, if we were to make an observation from a bio-computing aspect, it comes to light that bacteria possess a natural computing system \cite{key89}. In \cite{key84}, it has been suggested that behind the decision-making mechanisms of bacteria, there may be a hidden architecture resembling a neural-network. In \cite{key90}, they proposed using gene regulatory networks (GRNs) to perform artificial intelligence using molecular communication. Within each cell, distinctive gene regulatory networks operate, involving intricate interactions among regulatory proteins, RNA molecules, and specific DNA sequences. Bacteria use two-component systems to sense their environment. A sensor histidine kinase receives signals and a response regulator activates certain genes in these systems \cite{key110}. These two-component systems are viewed as a natural GRN pattern that can be modeled using ANN.

Artificial intelligence and machine learning studies are extremely important in situations where several species of bacteria coexist, such as the human gut. Given that these approaches excel in recognizing complex patterns within dynamic ecosystems, adapting to changes over time, and predicting interactions between species, they can significantly improve the efficiency and effectiveness of modeling and managing multi-species systems. In fact, automating this decision-making process can contribute to more precise and personalized models. In \cite{key90}, they discussed how multiple species of bacteria may be modeled using ML. If each bacterial species is considered as a point in the network, the links and edges between species can be formed by signaling molecules. Comparing this system with a generic feed-forward neural network, it can be seen that different bacterial species that receives the same signal can correspond to layers. Bacterial interactions and signal strength can be effected by factors such as intercellular communication, diversity of species, size of the populations, and how the signaling molecules diffuse throughout the system \cite{key90}. Larger populations receive and emit signals more quickly, behaving similarly to the weight of connections in ANN model. The implementation of such methods will facilitate the study of habitats inhabited by various bacterial species.

Although bacteria-based molecular machine learning is a promising field, there exist several challenges as listed below. 
\begin{itemize}
    \item \textbf{ML integration:} The integration of machine learning techniques with intricate and dynamic bacterial behavior, or bacterial colonies, is one of the most challenging aspects of this field. The constructed model must efficiently establish a connection between genetic codes and the resulting bacterial behavior, while also successfully navigating intricate environmental situations. Such models can be employed in scenarios with complex decision-making procedures, such as intelligent drug delivery systems \cite{key142}.
    \item \textbf{Automating the decision-making process:} Automating decision-making in bacteria through molecular machine learning requires creating synthetic genetic circuits and sensory systems in bacteria. This allows them to independently analyze environmental stimuli and make decisions using predetermined algorithms and feedback loops.
    \item \textbf{Datasets:} Machine learning applications rely heavily on high-quality and extensive datasets. Acquiring such data in the field of molecular communications and biological artificial intelligence is needed for applications such as improvement of plant or soil health or optimization circumstances for crop growth using  machine-learning-enhanced bacteria.
\end{itemize}

\subsection{Bacterial Integrated Sensing and Communications}
The evolution of the Internet of Things (IoT) and advancements in information and communication technologies has significantly broadened the sensing and communication capabilities of smart devices \cite{key94,key95}. This progress has played a pivotal role in shaping the development of integrated sensing and communication systems \cite{key91}. In broad terms, we can categorize systems as either primarily focused on sensing, primarily focused on communication, or a system designed through the co-design of both sensing and communication. In the context of joint design systems, the communication signal is anticipated to simultaneously serve the sensing function. Matrices such as the signal covariance matrix or precoding matrix can be employed in such designs \cite{key91}. Simultaneously, it is essential to design high-fidelity channel and target models to support the system \cite{key91}.

In this context, bacteria provide a suitable foundation for the integration of such systems due to some of their inherent properties. The natural attributes of bacteria make them conducive to integration of sensing and communication functionalities, aligning well with the requirements of these joint design systems. Integration offers several advantages, one of which is enhanced energy efficiency \cite{key92}. Bacteria, capable of generating energy through various metabolic processes, eliminate the need for external power sources. While current integrated systems often rely on radio frequency signals \cite{key93}, scenarios can be envisioned where the sensing system interacts with living organisms. In such applications, bacteria's notable advantage lies in their biocompatibility. Additionally, being biological entities, bacteria have a minimal environmental impact.

Research is still under progress in this field, but below are some of the necessary steps to achieve the integrated systems.

\begin{itemize}
    \item \textbf{Improving sensing capabilities:} Enhancing bacterial sensing abilities is necessary. Enhancing the precision and sensitivity of bacterial sensors is crucial for detecting a wider range of environmental signals and biomarkers.  
    \item \textbf{Robust communication systems:} In order to enhance the efficiency of transmitting information signals, it is necessary to enhance the dependability and range of bacteria in MC systems. 
    \item \textbf{Integration and data processing:} The sensed data must be incorporated into bacterial communication systems. The bacteria should activate a decision-making process based on the sensed inputs and use its communication skills accordingly. 
    \item \textbf{Microbial consortia:} It is necessary to create a structure where different bacteria are responsible for different sensing and communication tasks, just like in biofilms. In this way, a more efficient and advanced system can be created. 
    \item \textbf{Computational modeling:} Computational models should be developed to both predict and simulate the operation of these integrated sensing and communication systems, and the system should be designed and optimized accordingly. 
\end{itemize}

\subsection{Bacterial Quantum Computing}

The study of quantum mechanics and their applications in biology, as well as the field of quantum computing, have been gaining more traction recently. Quantum mechanics is a sub-discipline of physics that is generally based on the behaviour of atoms and subatomic particles in nature \cite{key111}. This extensively validated hypothesis has revolutionized our perspective on energy at the atomic scale. In general, there are some key quantum effects such as superposition \cite{key118}, entanglement \cite{key119}, interference \cite{key120}, coherence and tunneling \cite{key121}. Understanding these effects has resulted in the advancement of quantum technologies, such as quantum computing and quantum communication. 

Previously, it was thought that quantum mechanics could only explain some processes involving radiation for living organisms \cite{key112}. However, with the development of theoretical and experimental techniques, it has been seen that this assumption needs to be modified. Photosynthesis exemplifies the quantum phenomenon in biology through its process of light harvesting \cite{key113}. Photosynthetic pigment-protein complexes capture light and transfer the energy from the light as electronic excitation to the reaction center for the separation of charges. The utilization of sunlight relies on the transmission of energy and separation of charges, in which quantum phenomena like superposition and coherence have a significant impact. 

Green sulfur bacteria, as photosynthetic organisms, require light for survival \cite{key114}. The absorption of light and energy transmission to the reaction centers, where photosynthesis occurs, is facilitated by a specialized light-harvesting complex known as the chlorosome \cite{key114}. The chlorosome houses bacteriochlorophyll pigments, which are responsible for the absorption of light energy. However, their photosynthetic systems are susceptible to damage when exposed to oxygen. Consequently, the bacterium has evolved a mechanism to minimize this damage upon contact with oxygen. Researchers have identified this mechanism as vibronic mixing which is a quantum mechanical effect \cite{key113}. The phenomenon involves the combination of electrical states and vibrational states, leading to the formation of new hybrid states \cite{key115}. In other words, the vibrations mix harmoniously to the point where their individual characteristics become indistinguishable. Bacteria employ this process to channel energy towards locations where it will be most advantageous for their survival. Without oxygen present in the surroundings, energy is transferred ‘normally’ to the photosynthetic reaction center. However, when oxygen is present, the energy is suppressed and follows a less traditional pathway. This enables the system to adjust and maintain its integrity, albeit generating a reduced amount of energy compared to its regular output \cite{key113}.

\begin{figure*}[ht]
  \includegraphics[width=\linewidth]{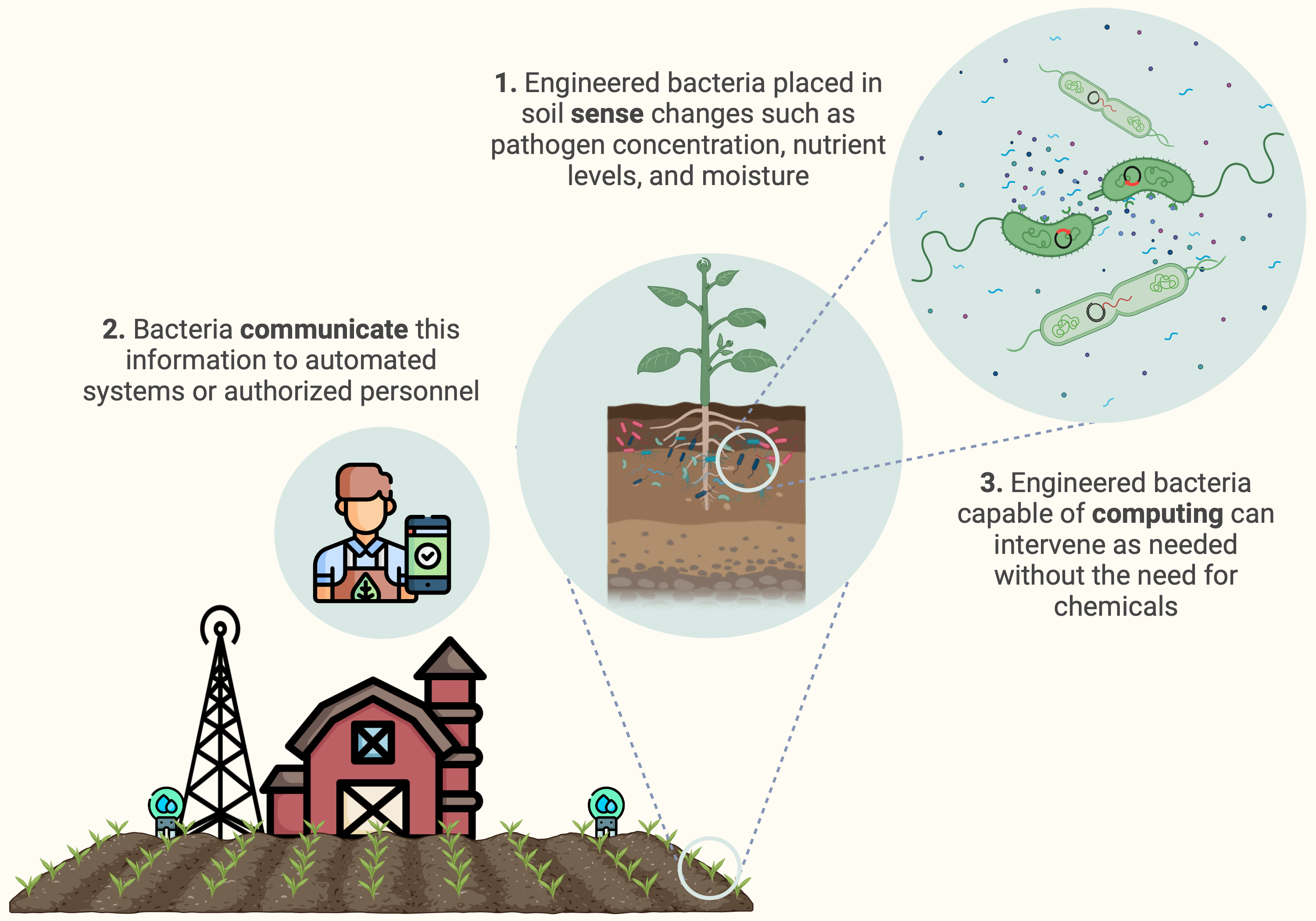}
    \caption{Integrated sensing, communication, and computing (ISCC) using bacteria for agricultural purposes. The genetically modified bacteria utilize their inherent sensing capabilities to detect and monitor many factors, such as the concentration of pathogens, nutrients, and moisture levels in the soil, as intended. The bacteria can share this information with other bacteria in the environment (conjugation) or communicate it to sensors or authorized personnel. Using the acquired characteristics that computing enables, bacteria can transform this information into useful data and facilitate the execution of appropriate activities, such as the production of certain chemicals or the elimination of pathogens. Bacteria are in a position to be one of the biggest enablers of IoE, especially for smart agricultural applications. Not only that, they may also have the potential to reduce or even eliminate the need for chemical use in agriculture. This will be a step towards solving some environmental problems, such as global warming, in the long run. (Created with BioRender.com)}
\end{figure*}

Moreover, in another study, researchers employed a completely quantized model to examine an experiment involving the introduction of green sulfur bacteria into a microcavity with clearly defined photon mode energies \cite{key116}. It was discovered that the bacteria's excitons and the photons in the cavity were strongly connected, leading to a change in the energy spectrum. This change introduced new energy levels known as polaritons. The system's ground state was determined to be an entangled state between the excitons in the bacterium and the photons in the cavity. This was observed through the phenomenon known as Rabi splitting \cite{key117}. This presents empirical support for the quantum characteristics of biological systems and introduces fresh opportunities for investigation in the field of quantum biology.

This behavior of bacteria has given rise to potential research fields. Although quantum computing has not yet achieved its peak, its undeniable potential holds significant promise. Some of the possible research directions are listed below.

\begin{itemize}
    \item \textbf{Quantum research on biological systems:} Further research is needed to better understand the effects of quantum phenomena on biological systems and to utilize these mechanisms in quantum computing.
    \item \textbf{Integration with artificial intelligence:} When combined with artificial intelligence, this field of study may hold a promising position in various industries, including drug development and antibiotic resistance. The capacity to accurately and rapidly model biochemical reactions and synthesis makes it a vital tool for improving research in these crucial areas.
    \item \textbf{Quantum devices and biological systems interface:} Interfaces that can integrate bacterial systems and traditional quantum computing devices should be developed. This might enable hybrid bio-quantum systems.  
    \item \textbf{Quantum aspects of electron transfer among bacteria:} Examining the quantum mechanical aspects of electron transfer among bacteria within the framework of nanowire-mediated communication and exploring the role of quantum effects in enhancing efficiency and specificity could establish a connection between quantum physics and biological systems.
\end{itemize}

\section{Future Prospects}

Thus far, we have enhanced our comprehension by delving into the intricate realm of bacteria, which may appear simple at first glance. The progress made so far has been characterized by notable advancements in MC, comprehension of bacterial behavior, and its implications in various situations. Nevertheless, despite these advancements, there remain still unresolved challenges as discussed in this paper. In addition, another innovative strategy we would like to propose is integrated sensing, communication and computing using bacteria. 

As previously stated, bacteria can be integrated into joint sensing and communication systems by utilizing their natural properties. Bacteria can be used in ways that go beyond their innate capabilities through engineering and the use of specific artificial intelligence methodologies \cite{key143, key144, key148, key145, key146, key147}. To enhance communication aspects, efforts have been directed towards addressing issues like communication rate and reliability. In this paper, several suggestions have been proposed to enhance the communication rate and reliability in the context of bacterial communication. Besides, advancements in synthetic biology are driving efforts to enhance the field of bacterial computing, both by attempting to adapt bacteria to existing models and by analyzing bacteria to create new models based on some of their innate advantages.

In light of these studies, developing integrated sensing, communication, and computing (ISCC) approaches using bacteria may lead to new research fields and open up new potential applications. Previously, ISCC systems for different purposes have been designed to enhance system performance and efficiency \cite{key130,key131, key212, key213}. However, the design of such a system incorporating bacteria is a new challenge. This integration utilizes bacterial systems as platforms for sensing, transmitting, and processing information at a molecular level. It involves using the inherent sensing and responsive skills of bacteria, in combination with engineered computing and communication functionalities, to develop systems capable of performing complex tasks.

The main aim here could be to create bacteria capable of sensing various information signals from the environment, using synthetic genetic circuits for examining these signals, and generating a targeted response accordingly. This integration might enable us to develop nano-devices that can be used in agriculture, biomedical applications, and energy harvesting systems.

ISCC has the potential to advance smart agriculture, which will probably be a must for us in the future, to a higher level. Engineered bacteria placed in the soil can detect changes such as pathogen concentration or nutrient levels and transmit this information to automated systems or authorized personnel. This, in turn, can trigger predetermined pathways for bacteria to intervene as needed. A representative figure of this system is shown in Figure 9. Similarly, bacteria can detect certain biomarkers in the human body at very early stages of the disease, communicate this information for diagnostic purposes, and compute the necessary responses, such as the production of therapeutic agents. In fact, with the involvement of artificial intelligence and machine learning, bacteria can even predict possible outcomes and offer different types of treatment methods.

The Internet of Things (IoT) has enabled us to connect various physical devices to the internet and create a network of objects that are embedded with sensors, software, and other technologies \cite{key95, key216, key217, key218, key219}. On the other hand, Internet of Everything (IoE) goes a step further and offers ubiquitous connectivity by also involving people, processes, and data \cite{key37,key214,key215}. In simple terms, IoE seeks to connect all aspects of our world through the internet to create more efficient systems and improve decision-making. This proposed approach of integrating sensing, communication, and computing using bacteria has the potential to be one of the biggest enabler of IoE for various applications.

\section{Conclusion}

In this paper, we summarised the communication networks of bacteria, which can be used both as an internal communication and as an information-carrying signal. We aimed to shed light on molecular communication studies involving the use of bacteria, particularly by highlighting some fundamental properties that bacteria should possess in communication networks. Furthermore, we have delved into bacterial computing studies, assessing them not solely in the context of communication aspects but also exploring their potential areas of study within computing.

Our aim was to review bacterial communication and computing from the perspective of molecular communication, address relevant issues inherent to this viewpoint, highlight certain models in the literature, and provide insights into potential pathways for future studies. In addition to exploring bacterial communication methods, we have sought to unveil the vast potential of bacteria for the future by delving into various areas, ranging from the sensing mechanisms they employ to the roles they play in computing and communication systems as carriers of information. Hence, we have charted a roadmap that begins with quorum sensing and extends towards quantum computing.

We hope to inspire follow-up work in two directions. On the one hand, the provided review can guide researchers who want to embark on studying molecular communication, especially using bacteria. On the other hand, our insights can assist the community in developing new models and integrated systems using bacteria, which are necessary for future applications.

\ifCLASSOPTIONcaptionsoff
  \newpage
\fi
\bibliographystyle{IEEEtran}
\bibliography{bacterialcomm}
\end{document}